\documentclass[11pt,a4paper]{article}

\usepackage{amsmath,amssymb,tikz,hyperref,empheq}
\usepackage{graphicx}
 \usepackage{verbatim}

\usepackage{tikz}
\usetikzlibrary {decorations.pathmorphing}
\usetikzlibrary {arrows.meta}
\usetikzlibrary{decorations.text}
 \allowdisplaybreaks

\def\be{\begin{equation}}
\def\ee{\end{equation}}
\def\ba#1\ea{\begin{align}#1\end{align}}
\def\bg#1\eg{\begin{gather}#1\end{gather}}
\def\bm#1\em{\begin{multline}#1\end{multline}}
\def\bmd#1\emd{\begin{multlined}#1\end{multlined}}

\def\({\left(}
\def\){\right)}
\def\[{\left[}
\def\]{\right]}

\def \be {\begin{equation}}
\def \ee {\end{equation}}
\def \ba {\begin{array}}
\def \ea {\end{array}}
\def \bea{\begin{eqnarray}}
\def \eea{\end{eqnarray}}

\def\bea{\begin{eqnarray}}
\def\eea{\end{eqnarray}}

\newcommand{\bit}{\begin{itemize}}  \newcommand{\eit}{\end{itemize}}
\newcommand{\ben}{\begin{enumerate}}  \newcommand{\een}{\end{enumerate}}

\long\def\symbolfootnote[#1]#2{\begingroup%
\def\thefootnote{\fnsymbol{footnote}}\footnote[#1]{#2}\endgroup}

\newcommand{\sysu}{{\it School of Physics and Astronomy, Sun Yat-Sen University, 2 Daxue Road, Zhuhai 519082, China}}

\begin{document}
\thispagestyle{empty}
\begin{center}

~\vspace{20pt}

{\Large\bf Entanglement Island and Page Curve in Wedge Holography}

\vspace{25pt}

Rong-Xin Miao ${}$\symbolfootnote[1]{Email:~\sf
  miaorx@mail.sysu.edu.cn}

\vspace{10pt}${}$\sysu

\vspace{2cm}

\begin{abstract}
Entanglement islands play an essential role in the recent breakthrough in resolving the black hole information paradox. However, whether entanglement islands can exist in massless gravity theories is controversial. It is found that entanglement islands disappear in the initial model of wedge holography with massless gravity on the brane. As a result, the entanglement entropy of Hawking radiation becomes a time-independent constant, and there is no Page curve. In this paper, we recover massless entanglement islands in wedge holography with suitable DGP gravity or higher derivative gravity on the branes. We study two typical cases. In the first case, we consider a black hole on the strong-gravity brane and a bath on the weak-gravity brane. It is similar to the usual double holography with non-gravitational baths. In the second case, we discuss two black holes on the two branes with the same gravitational strength. We recover massless entanglement islands and non-trivial Page curves in both cases. We also argue that the entanglement island is consistent with massless gravity. Our results strongly support that entanglement islands can exist in long-range theories of gravity.
\end{abstract}

\end{center}

\newpage
\setcounter{footnote}{0}
\setcounter{page}{1}

\tableofcontents

\section{Introduction}

Recently, there has been a significant breakthrough toward resolving the black hole information paradox \cite{Hawking:1976ra}, where the entanglement islands play a critical role \cite{Penington:2019npb,Almheiri:2019psf,Almheiri:2019hni}. See \cite{Almheiri:2020cfm} for a good review. For simplicity, one considers the Hawking radiation emitted into a non-gravitational bath. This can be naturally realized in doubly holographic models such as Karch-Randall (KR) braneworld \cite{Karch:2000ct} and AdS/BCFT \cite{Takayanagi:2011zk}. 
\begin{figure}[t]
\centering
\includegraphics[width=9cm]{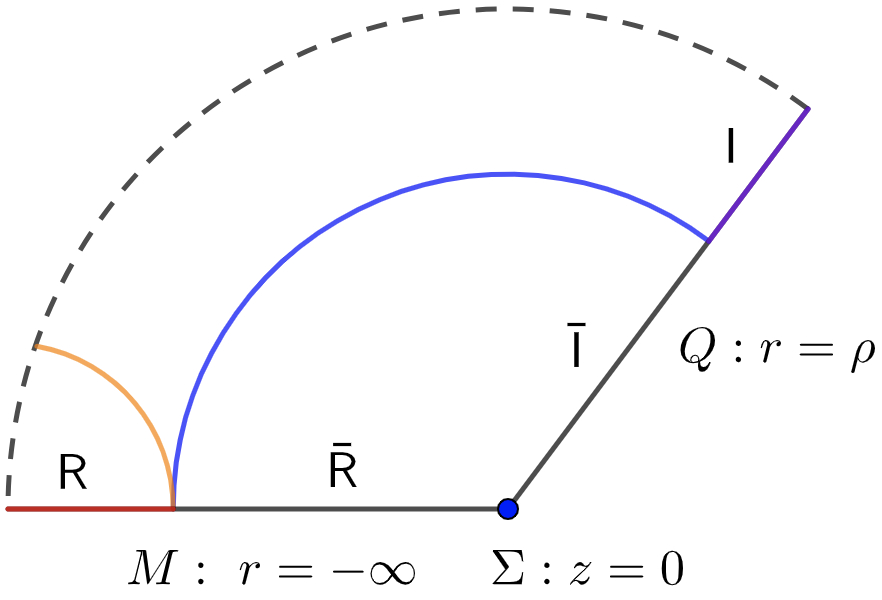}
\caption{ The black-string geometry and its interpretation in black hole information paradox. $Q$ is the KR brane with a gravitational black hole, and $M$ is the AdS boundary with a non-gravitational black hole (bath). $\text{I}$ and $\bar{\text{I}}$ denotes the island region (purple line) and its complement (black line) on the brane $Q$, $\text{R}$ and $\bar{\text{R}}$ denotes the radiation region (red line) and its complement (black line) on the AdS boundary $M$, $\Sigma$ is the defect (blue point) on the corner. Note that the island region $\text{I}$ and radiation region $\text{R}$ envelop the black-hole horizon on $Q$ and $M$, respectively. For simplicity, we only show the regions outside the horizon. In bulk, the dotted line, blue, and orange lines indicate the horizon, RT surfaces in the island phase, and Hartman-Maldacena (HM) surface in the no-island phase at $t=0$, respectively.}
\label{doubleholography}
\end{figure}
Let us take the doubly holographic black-string model \cite{Geng:2021mic} as an example
 \begin{eqnarray}\label{blackstring}
ds^2=dr^2+\cosh^2(r) \frac{\frac{dz^2}{f(z)}-f(z)dt^2+\sum_{\hat{i}=1}^{d-2}dy_{\hat{i}}^2}{z^2},
\end{eqnarray}
where $f(z)=1-z^{d-1}$ with the horizon at $z=1$, $r$ denotes the distance to the brane, the Karch-Randall (KR) brane $Q$ locates at $r=\rho$, and the AdS boundary $M$ is at $r=-\infty$. See Fig.\ref{doubleholography} for the geometry, where a gravitational black hole lives on the KR brane $Q$, and a non-gravitational black hole (bath) locates on the AdS boundary $M$ \footnote{Note that there is a black hole on the AdS boundary for the black string, which is different from the usual double holography with AdS-Schwarzschild-like black holes in bulk. }.  One imposes the transparent boundary condition on the defect $\Sigma$ so that Hawking radiation on $Q$ can flow into the bath on $M$. It proposes that one should use the following island rule to calculate the entanglement entropy of Hawking radiation R
\begin{eqnarray}\label{islandrule}
S_{\text{EE}}(\text{R})=\text{min} \Big\{ \text{ext} \Big( S_{\text{QFT}}(\text{R}\cup \text{I})+ \frac{A(\partial \text{I})}{4 \hat{G}_N }\Big) \Big\},
\end{eqnarray}
where one adjusts the island region $\text{I}$ to minimize the above generalized entropy \cite{Faulkner:2013ana,Engelhardt:2014gca} . It is believed that one can extract information on the island from the radiation region R, although they are disconnected in the lower-dimensional system $Q\cup M$. Interestingly, the entanglement entropy of QFT can decrease by adding a disconnected region, i.e., $S_{\text{QFT}}(\text{R}\cup \text{I})< S_{\text{QFT}}(\text{R})$. This quantum property is important in reducing the entropy and recovering the Page curve. So far, exact derivations of the Page curve are limited to Jackiw–Teitelboim gravity in two dimensions, where there are no gravitons. In higher dimensions, all reliable discussions focus on doubly holographic models. See \cite{Almheiri:2019psy,Geng:2020qvw,Chen:2020uac,Ling:2020laa} for examples.  See also
\cite{Rozali:2019day,Chen:2019uhq,Almheiri:2019yqk,Kusuki:2019hcg,
   Balasubramanian:2020hfs,  Kawabata:2021hac,Bhattacharya:2021jrn,Kawabata:2021vyo,Chen:2020hmv,Krishnan:2020fer,Ghosh:2021axl,Bhattacharya:2021nqj,Geng:2021mic,Chou:2021boq,Ahn:2021chg,Alishahiha:2020qza,Gan:2022jay,Omidi:2021opl,Hu:2022ymx,Azarnia:2021uch,Anous:2022wqh,Saha:2021ohr,Yadav:2022mnv,Geng:2022slq,Geng:2022tfc,Yu:2022xlh,Chu:2022ieq,Hu:2022zgy} for some recent works on entanglement islands and Page curve.

Unfortunately, the gravity on the brane is massive in the usual double holography such as Karch-Randall (KR) braneworld \cite{Karch:2000ct} and AdS/BCFT \cite{Takayanagi:2011zk} \footnote{See also \cite{Miao:2017gyt,Chu:2017aab,Miao:2018qkc,Chu:2021mvq} for other proposals of AdS/BCFT with various boundary conditions.}. Physically, that is because a gravitational system on the brane $Q$ is coupled with a non-gravitational system on the AdS boundary $M$. As a result, the general covariance breakdowns, leading to a mass for the graviton. Technically, that is because one imposes Neumann boundary condition (NBC) on the brane $Q$ while imposing Dirichlet boundary condition (DBC) on the AdS boundary $M$. However, according to \cite{Hu:2022lxl}, only if one sets both NBCs on the two boundaries $Q$ and $M$ does the massless gravity appear. Besides, the massless mode is non-normalizable since the AdS boundary $M$ locates at infinity. 

\begin{figure}[t]
\centering
\includegraphics[width=9cm]{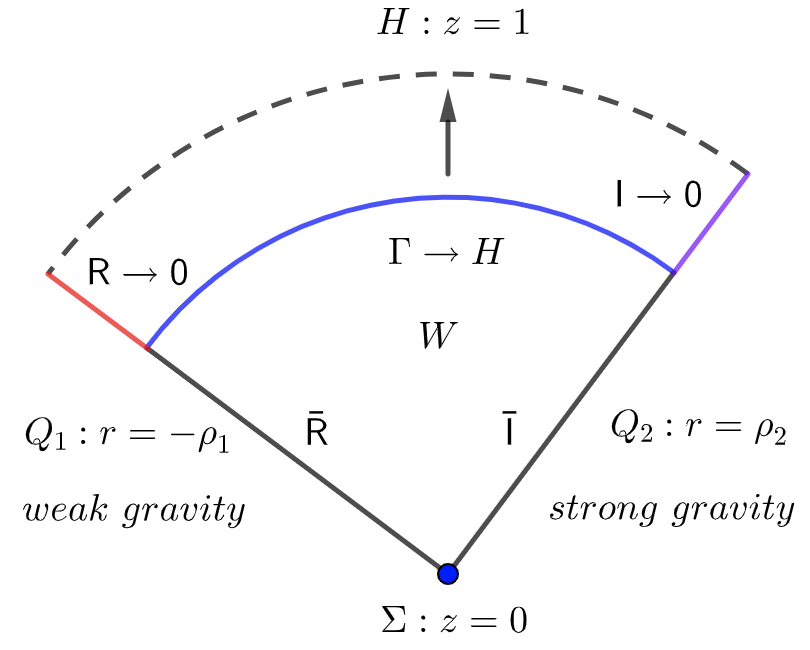}
\caption{The geometry of wedge holography without DGP terms and its interpretation in black hole information paradox. $W$ is the bulk wedge space, $Q_1$ is the weak-gravity ``bath" brane and $Q_2$ is the strong-gravity brane, $\Sigma$ is the defect on the corner of the wedge, $H$ denotes the horizon in bulk. According to  \cite{Geng:2020fxl}, since both branes are gravitating, one should adjust both the radiation region $\text{R}$ (red line) and the island region $\text{I}$ (purple line) to minimize the entanglement entropy in the island phase. Remarkably, the corresponding RT surface $\Gamma$ (blue line) coincides with the horizon $H$ (black dotted line). As a result, the potential island and radiation regions I and R of Fig. \ref{doubleholography} disappear. Note that the island region envelops the black-hole horizon on the brane $Q_2$, and only the region outside the horizon disappears.}
\label{wedge without DGP}
\end{figure}

Naturally, we get normalizable massless gravity if we set $M$ at a finite place as $Q$ and impose both NBCs on the two boundaries. This deformed double holography is called wedge holography \cite{Akal:2020wfl, Miao:2020oey} \footnote {The original motivation of wedge holography is not to obtain massless gravity. The existence of massless gravity in wedge holography is found in \cite{Hu:2022lxl}. See also \cite{Geng:2020fxl}}. See Fig. \ref{wedge without DGP} for the geometry and \cite{Miao:2021ual} for its generalization to codim-n defects. Wedge holography proposes that the classical gravity in the $(d+1)$-dimensional bulk $W$ is dual to ``quantum gravity" on the $d$-dimensional branes $Q=Q_1\cup Q_2$ and is dual to the conformal field theory (CFT) on the $(d-1)$-dimensional corner $\Sigma$. Thus, it is also called codim-2 holography. In wedge holography, the effective theory on the branes is a CFT plus a ghost-free higher derivative gravity, or an equivalent multi-metric gravity, which behaves like Einstein gravity in many aspects \cite{Hu:2022lxl}. For example, they yield the same holographic Weyl anomaly and the first law of entanglement entropy. Besides, all of the solutions to Einstein gravity are also solutions to the effective higher derivative gravity on the branes.

\begin{figure}[t]
\centering
\includegraphics[width=6.5cm]{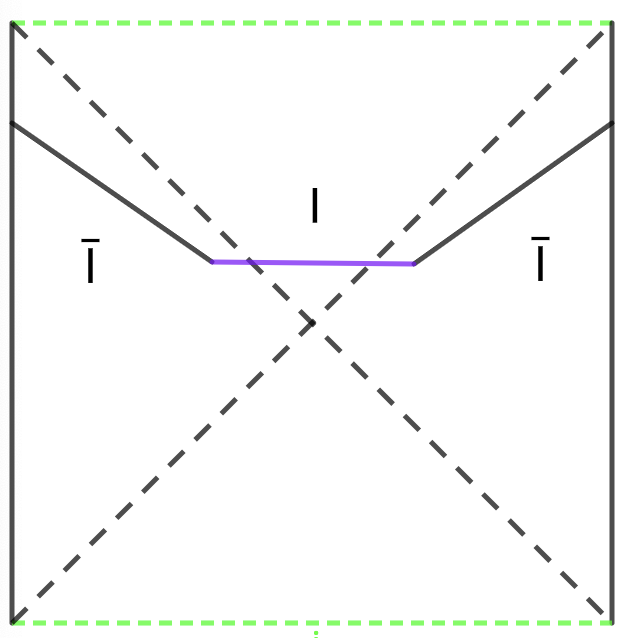}
\includegraphics[width=6.5cm]{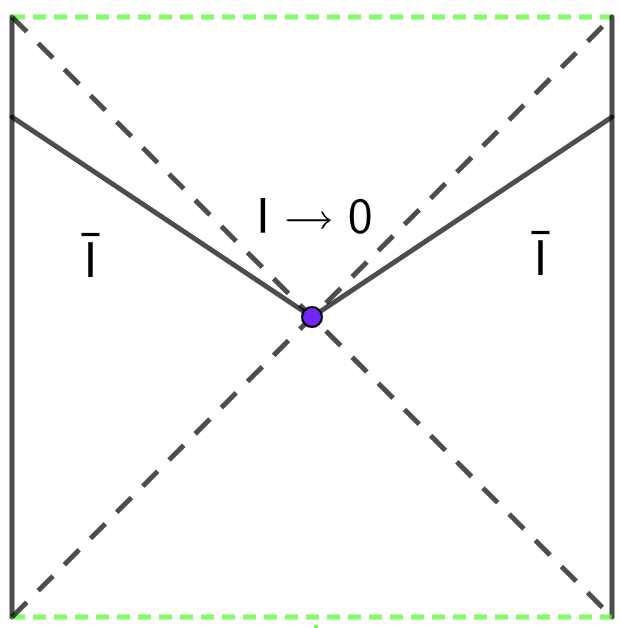}
\caption{ Left: Penrose diagram on brane $Q$ in the usual double holography (Fig.\ref{doubleholography}). Right: Penrose diagram on the strong-gravity brane $Q_2$ in wedge holography (Fig.\ref{wedge without DGP}). The black-dotted line, green-dotted line, and the purple line or point denote the horizon, singularity and island, respectively. It shows that the island shrinks into a point in the Penrose diagram of wedge holography. In this sense, it claims that the entanglement island disappears in wedge holography.}
\label{Penrose diagram with and without island}
\end{figure}

Unfortunately, although we have massless gravity in wedge holography, the entanglement island disappears \cite{Geng:2020fxl, Geng:2021hlu}.  
Let us explain how this happens in Fig. \ref{wedge without DGP}.  According to \cite{Geng:2020fxl}, since both branes are gravitating in wedge holography, one should adjust both the radiation region $\text{R}$ and the island region $\text{I}$ to minimize the entanglement entropy of Hawking radiation.  Moreover, from the viewpoint of bulk, since the RT surface is minimal, it is natural to adjust its intersections $\partial \text{R}$ and $\partial \text{I}$ on the two branes to minimize its area. 
Following this approach, the RT surface (blue line) in the island phase coincides with the horizon (dotted line)  \cite{Geng:2020fxl}. As a result, the island and radiation regions I and R of Fig. \ref{doubleholography} disappear, and the entanglement entropy of radiation emitted into the bath becomes a time-independent constant \cite{Geng:2020fxl}. Note that the island region (purple line) envelops the black-hole horizon on the brane $Q_2$, and only the region outside the horizon disappears. See also the Penrose diagram in Fig.(\ref{Penrose diagram with and without island}) (right), which shows that the island shrinks into a point in wedge holography. In this sense, we say that the entanglement island disappears in wedge holography \footnote{ The ``island" is a broad concept in the literature. Here
the island disappears in the sense of the Penrose diagram as shown in Fig.(\ref{Penrose diagram with and without island}) (right). On the other hand, if one defines the island as RT surfaces ending on the branes \cite{Chen:2020uac}, of course, there is an island in that sense. The critical point here is that the entanglement entropy of Hawking radiation is a time-independent constant, and there is no Page curve in wedge holography when there is massless gravity on the branes \cite{Geng:2020fxl, Geng:2022fui}.}. Inspired by the above observation, it is conjectured that the entanglement island can exist only in massive gravity theories \cite{Geng:2021hlu, Geng:2022fui}. They argue that the entanglement island is inconsistent with massless gravity obeying Gauss's law. There are controversies on this conjecture \cite{Krishnan:2020fer, Ghosh:2021axl}. According to \cite{Almheiri:2020cfm}, it is natural that the island mechanism works for massless gravity. Interestingly, \cite{Emparan:2023dxm} finds that the absence-of-island issue can be ameliorated in the large D limit.

It is a significant question whether there are entanglement islands in massless gravity. From a practical point of view, the gravity in our universe is massless. Strict experimental limits on gravity mass have also been set based on the gravitational wave, Yukawa potential, dispersion relation, and modified gravity theories \cite{LIGOScientific:2016lio, Experimental limits gravity mass}. Thus, addressing the information paradox in the real world is more crucial and necessary than in a toy model of massive gravity. From a theoretical point of view, massive gravity suffers the non-causal problem. Although the ghost-free theory can be constructed \cite{deRham:2010kj}, massive gravity admits superluminal shock wave solutions and thus violates causality generally \cite{Deser:2012qx}. It is not satisfactory if the island rule applies only to an acausal theory. This paper gives a positive answer to the above question. We find that massless entanglement islands can exist in wedge holography with Dvali-Gabadadze-Porrati (DGP) gravity \cite{Dvali:2000hr} or higher derivative gravity on the branes. It helps to clarify the theoretical controversy and strongly implies that the entanglement island is consistent with massless gravity theories.

This paper investigates many aspects of wedge holography with DGP terms. We find that there is normalizable massless gravity on the branes. By analyzing effective Newton's constants, brane bending modes, and holographic entanglement entropy, we obtain several lower bounds for the DGP parameters. Interestingly, the DGP parameters can be negative. We discuss the Page curve for eternal two-side black holes in this paper. For simplicity, we show only one side of the systems in most figures (Fig.\ref{doubleholography}, Fig.\ref{wedge without DGP}, Fig.\ref{Wedge1}, Fig.\ref{Wedge2}). One can double these figures for the two-side geometry as in Figure 1 of \cite{Geng:2021mic}. We discuss two different situations. In case I shown in Fig.\ref{Wedge1}, we take approximately the black hole on weak-gravity brane $Q_1$ as the ``bath" and focus on the Hawking radiation of the black hole on the strong-gravity brane $Q_2$. Following \cite{Geng:2020fxl}, the primary purpose of this approximation is to mimic the usual case with a non-gravitating bath. We call it ``case I: one black hole approximately" in sect. 3.
One may ask what happens if we take the two black holes on $Q_1\cup Q_2$ seriously. This is the motivation we further consider case II, shown in Fig.\ref{Wedge2}. The two branes have equal gravitational strength in case II. Thus, there is no natural way to choose which black hole is the ``bath" black hole, and we name it ``case II: two black holes" in sect. 4. We recover massless entanglement islands and Page curves in both cases. We argue that the entanglement islands can consistently exist in the brane-world models of massless gravity. Finally, we generalize the results to higher derivative gravity on the branes. 

The paper is organized as follows. 
In section 2, we formulate wedge holography with DGP gravity on the brane. Then, we show massless gravity on the branes and get several lower bounds for the DGP parameter. Section 3 discusses the entanglement island and the Page curve in case I: one strong-gravity black hole coupled with a weak-gravity bath black hole. Section 4 generalizes the discussions to case II: two black holes associated with two strong-gravity baths. Section 5 discusses the possible resolutions to the puzzle of the massless island raised by \cite{Geng:2021hlu}. Section 6 generalizes the discussions to higher derivative gravity on the branes. Finally, we conclude with some open problems in section 7.  

Note that parts of the results have been shown in the letter \cite{Miao:2022mdx}. We give more details and new developments in this paper. The new results include the mass spectrum, brane bending mode, holographic entanglement entropy, details for calculations of Page curves, an inspiring analog of the island puzzle and its possible resolutions in AdS/CFT, and generalizations to higher derivative gravity on the branes.

\section{Wedge holography with DGP terms}

This section investigates the wedge holography with DGP gravity on the brane. First, we work out the effective action for one novel class of solutions and verify that there is normalizable massless gravity on the brane. We get a lower bound of the DGP parameter to have a positive effective Newton's constant. Second, we find the mass spectrum on the brane obeys the Breitenlohner-Freedman bound $m^2\ge -(d-1)^2/4$, so the system is tachyon-free. Third, we derive the effective action of brane bending modes, which yields another lower bound of the DGP parameter. 
Finally, we discuss the holographic entanglement entropy and get an additional lower bound of the DGP parameter. 

Let us recall the geometry of wedge holography shown in Fig.\ref{wedge without DGP}, where $W$ is the bulk wedge space, $Q=Q_1\cup Q_2$ denote two end-of-the-world branes, $\Sigma$ labels the corner of the wedge, where the defect lives. Let us take a typical metric to illustrate the geometry 
\begin{eqnarray}\label{AdSmetric}
ds^2=dr^2+\cosh^2(r)\frac{dz^2-dt^2+\sum_{\hat{i}=1}^{d-2} dy^2_{\hat{i}}}{z^2}, \ -\rho_1 \le r\le \rho_2,
\end{eqnarray}
where the left brane $Q_1$, the right brane $Q_2$, and the defect $\Sigma$ locate at $r=-\rho_1$, $r=\rho_2$ and $z=0$, respectively. Wedge holography has three equivalent descriptions:

1. a classical gravity coupled with two branes in the $(d+1)$-dimensional bulk $W$,

2. a ``quantum gravity" coupled with CFTs on the $d$-dimensional branes $Q=Q_1\cup Q_2$,

3. a CFT on the $(d-1)$-dimensional defect $\Sigma$.

Now we quickly formulate wedge holography with DGP gravity on the branes. The action is given by
\begin{eqnarray}\label{action}
I=\int_W dx^{d+1}\sqrt{-g}\Big(R_W+d(d-1)\Big)+2\int_{Q} dx^d\sqrt{-h_Q}(K-T_a+\lambda_a R_{Q}),
\end{eqnarray}
where $R_W$ is the Ricci scalar in bulk $W$, $K$ is the extrinsic curvature, $h_{Q\ ij}$ and $R_{Q}$ are the induced metric and the intrinsic Ricci scalar (DGP term) on the branes $Q=Q_1\cup Q_2$, and $T_a$ and $\lambda_a$ with $a=1,2$ are free parameters. For simplicity, we have set Newton's constant $16\pi G_N=1$ together with the AdS radius $L=1$. 
Following \cite{Takayanagi:2011zk}, we choose Neumann boundary condition (NBC) on $Q$ 
\begin{eqnarray}\label{NBC}
K^{ij}-(K-T_a+\lambda_a R_{Q}) h_Q^{ij}+2 \lambda_a R^{ij}_Q=0,
\end{eqnarray}
which yields a massless gravitational mode on the brane \cite{Hu:2022lxl}. On the other hand, the gravity becomes massive if one imposes Dirichlet boundary condition (DBC) \cite{Miao:2018qkc} or conformal boundary condition (CBC) \cite{Chu:2021mvq} on one or two of the branes. In general, it isn't easy to find exact solutions which satisfy both Einstein equations in bulk and NBC (\ref{NBC}) on the boundary. 

\subsection{Effective action}

Fortunately, there is one novel class of exact solutions
\cite{Miao:2020oey}
\begin{eqnarray}\label{metric}
ds^2=dr^2+\cosh^2(r) h_{ij}(y) dy^i dy^j, \ -\rho_1 \le r\le \rho_2,
\end{eqnarray}
if $h_{ij}$ obeys Einstein equations on the brane
\begin{eqnarray}\label{EinsteinEQ}
 R_{h\ ij}-\frac{R_h+(d-1)(d-2)}{2} h_{ij}=0,
 \end{eqnarray} 
 and the brane tensions are given by
 \begin{eqnarray}\label{Ta}
T_a=(d-1) \tanh(\rho_a)-\lambda_a \frac{(d-1)(d-2)}{\cosh^2(\rho_a)}.
 \end{eqnarray} 
Note that $R_h$ of (\ref{EinsteinEQ}) denotes Ricci scalar defined by $h_{ij}$, which is different from $R_Q$ defined by $h_{Q\ ij}=\cosh(\rho_a) h_{ij}$.  
 Substituting the metric (\ref{metric}) into the action (\ref{action}) and integrating $r$, we get an effective action on each brane
  \begin{eqnarray}\label{effective action}
I_a=\frac{1}{16\pi G^a_{\text{eff N}}}\int_{Q_a} \sqrt{-h} \Big( R_h+(d-1)(d-2) \Big),
 \end{eqnarray} 
where $R_h$ is the Ricci scalar defined by $h_{ij}$ and $G^a_{\text{eff N}}$ denotes the effective Newton's constant
  \begin{eqnarray}\label{effective Newton constant}
\frac{1}{16\pi G^a_{\text{eff N}}}=2\lambda_a \cosh^{d-2}(\rho_a)+\int_0^{\rho_a} \cosh^{d-2}(r) dr.
 \end{eqnarray} 
In the above derivations, we have used (\ref{Ta}), $K=d \tanh\rho_a$, $h_{Q\ ij}=\cosh^2(\rho_a) h_{ij}$, $R_Q= \text{sech}^2(
\rho_a) R_h$ and
\begin{eqnarray}\label{Rh}
R_W=R_{h} \text{sech}^2(r)- d\left( 2+(d-1) \tanh^2r\right).
\end{eqnarray}

\begin{figure}[t]
\centering
\includegraphics[width=9cm]{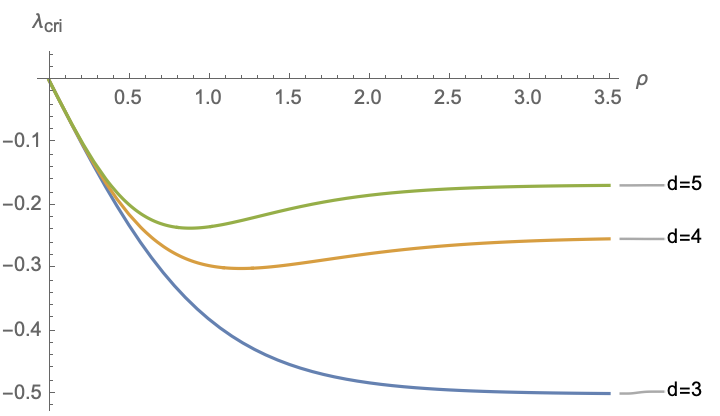}
\caption{ The lower bound $ \lambda_{\text{cri}}(\rho)$ in various dimensions. The larger the spacetime dimension is, the larger the lower bound is.}
\label{lambdabound1}
\end{figure}

From the EOM (\ref{EinsteinEQ}) and effective action (\ref{effective action}), it is clear that there is massless gravity on the branes. We require that the effective Newton's constant (\ref{effective Newton constant}) is positive, which yields a lower bound on the DGP parameter
\begin{eqnarray}\label{boundDGP1}
\lambda_a \ge \lambda_{\text{cri}}(\rho_a)=-\frac{1}{2} \int_0^{\rho_a} \frac{\cosh^{d-2}(r)}{ \cosh^{d-2}(\rho_a)} dr.
\end{eqnarray}
We draw $ \lambda_{\text{cri}}(\rho)$ in Fig.\ref{lambdabound1}, which shows that $ \lambda_{\text{cri}}$ has a lower bound too
\begin{eqnarray}\label{boundDGP1a}
 \lambda_{\text{cri}}(\rho_a) \gtrsim \begin{cases}
-\frac{1}{2} ,
 \ \ \ \ \ \ \text{for}\ d=3,\\
-0.300,
\ \text{for} \ d=4,\\
-0.236,
\ \text{for} \ d=5.
\end{cases}
\end{eqnarray}
The larger the spacetime dimension is, the larger the lower bound is. In the large $d$ limit, $\lambda_{\text{cri}}$ approaches zero, i.e., $\lim_{d\to \infty}\lambda_{\text{cri}}\to 0$.

\subsection{Mass spectrum}

In this subsection, we study the mass spectrum of gravitons on the branes in wedge holography with DGP terms. We find the mass spectrum obeys Breitenlohner-Freedman bound $m^2\ge -(d-1)^2/4$. In particular, it includes a massless mode, which agrees with the results of the last subsection. We focus on the fixed brane locations in this subsection, which yield $m^2\ge 0$. We leave the discussions of brane bending modes with $m^2=-(d-2)$ to the following subsection. 
 
We take the following ansatz of the perturbation metric and the embedding function of $Q$
 \begin{eqnarray}\label{perturbationmetric}
&&ds^2=dr^2+\cosh^2 (r) \left( h^{(0)}_{ij}(y) + \epsilon H(r) h^{(1)}_{ij}(y)  \right)dy^i dy^j+O(\epsilon^2),\\
&& Q_1: \ r=-\rho_1+O(\epsilon^2), \ \ \ \ \ Q_2: r=\rho_2+O(\epsilon^2), \label{perturbationQ1}
\end{eqnarray}
where $h^{(0)}_{ij}(y)$ is the AdS metric with a unit radius and $h^{(1)}_{ij}(y)$ denotes the perturbation, $\epsilon$ denotes the order of perturbations.  
In terms of bulk metric perturbations,  we have
 \begin{eqnarray}\label{bulkmetricperturbations}
\delta g_{r\mu}=0,\ \delta g_{ij}=\cosh^2 (r)  H(r) \bar{h}^{(1)}_{ij}(y).
\end{eqnarray}
Imposing the transverse traceless gauge 
 \begin{eqnarray}\label{gijgauge}
\nabla^{\mu} \delta g_{\mu\nu}=0,\ \ \  g^{\mu\nu}\delta g_{\mu\nu}=0,
\end{eqnarray}
we get
 \begin{eqnarray}\label{hij1gauge}
D^i h^{(1)}_{ij}=0,\ \ \  h^{(0)ij}h^{(1)}_{ij}=0,
\end{eqnarray}
where $\nabla_{\mu}$ and $D_i$ are the covariant derivatives with respect to $g_{\mu\nu}$ and $h^{(0)}_{ij}$, respectively.  Substituting (\ref{perturbationmetric}) and (\ref{hij1gauge}) into Einstein equations and separating variables, we obtain
 \begin{eqnarray}\label{EOMMBCmassivehij}
&& \left(\Box+2-m^2\right)h^{(1)}_{ij}(y)=0,\\
&& \cosh^2(r) H''(r)+d \sinh (r) \cosh (r) H'(r) + m^2 H(r)=0, \label{EOMMBCmassiveH}
\end{eqnarray}
where $m$ denotes the mass of gravitons and $\Box=D_k D^k$ is the d'Alembert operator defined by $h^{(0)}_{ij}$. Solving (\ref{EOMMBCmassiveH}), we derive
 \begin{eqnarray}\label{massiveHsolution}
H(r)=\text{sech}^{\frac{d}{2}}(r) \left(c_1 P_{\lambda_g}^{\frac{d}{2}}(\tanh r)+c_2 Q_{\lambda_g}^{\frac{d}{2}}(\tanh r)\right),
\end{eqnarray} 
where $P_{\lambda_g}^{\frac{d}{2}}$ and $ Q_{\lambda_g}^{\frac{d}{2}}$ are the Legendre polynomials, $c_1$ and $c_2$ are integral constants and $\lambda_g$ is given by
 \begin{eqnarray}\label{aibia1}
\lambda_g=\frac{1}{2} \left(\sqrt{(d-1)^2+4  m^2}-1\right),
\end{eqnarray}
which yields the correct Breitenlohner-Freedman bound of massive gravity in $\text{AdS}_d$
 \begin{eqnarray}\label{BFgravity}
 m^2\ge -(\frac{d-1}{2})^2.
\end{eqnarray}

By using EOM (\ref{EOMMBCmassivehij}), we can simplify the NBC (\ref{NBC}) as
 \begin{eqnarray}\label{NBCH1}
&&\cosh ^2\left(\rho _1\right) H'\left(-\rho _1\right)+2 \lambda _1 m^2 H\left(-\rho _1\right)=0,\\
&&\cosh ^2\left(\rho _2\right) H'\left(\rho _2\right)-2 \lambda _2 m^2 H\left(\rho _2\right)=0. \label{NBCH2}
\end{eqnarray}
Substituting the solution (\ref{massiveHsolution}) into (\ref{NBCH1},\ref{NBCH2}), we derive a constraint for the mass
 \begin{eqnarray}\label{spectrum}
m^2\big( M_{00} + M_{10} \lambda_1+ M_{01} \lambda_2+ M_{11} \lambda_1  \lambda_2 \big)=0,
\end{eqnarray}
with
 \begin{eqnarray}\label{M00}
&&M_{00}=\sqrt{1-x_1^2} \sqrt{1-x_2^2} \left(P_{\lambda _g}^{\frac{d}{2}-1}\left(x_2\right) Q_{\lambda _g}^{\frac{d}{2}-1}\left(-x_1\right)-P_{\lambda _g}^{\frac{d}{2}-1}\left(-x_1\right) Q_{\lambda _g}^{\frac{d}{2}-1}\left(x_2\right)\right),\\ \label{M10}
&&M_{10}=2 \left(x_1^2-1\right) \sqrt{1-x_2^2} \left(P_{\lambda _g}^{\frac{d}{2}}\left(-x_1\right) Q_{\lambda _g}^{\frac{d}{2}-1}\left(x_2\right)-P_{\lambda _g}^{\frac{d}{2}-1}\left(x_2\right) Q_{\lambda _g}^{\frac{d}{2}}\left(-x_1\right)\right),\\ \label{M01}
&&M_{01}=2 \sqrt{1-x_1^2} \left(x_2^2-1\right) \left(P_{\lambda _g}^{\frac{d}{2}}\left(x_2\right) Q_{\lambda _g}^{\frac{d}{2}-1}\left(-x_1\right)-P_{\lambda _g}^{\frac{d}{2}-1}\left(-x_1\right) Q_{\lambda _g}^{\frac{d}{2}}\left(x_2\right)\right),\\
&&M_{11}=-4 \left(x_1^2-1\right) \left(x_2^2-1\right) \left(P_{\lambda _g}^{\frac{d}{2}}\left(x_2\right) Q_{\lambda _g}^{\frac{d}{2}}\left(-x_1\right)-P_{\lambda _g}^{\frac{d}{2}}\left(-x_1\right) Q_{\lambda _g}^{\frac{d}{2}}\left(x_2\right)\right),\label{M11}
\end{eqnarray}
where $x_1=\tanh\rho_1, x_2=\tanh\rho_2$ and $\lambda_g$ is given by (\ref{aibia1}).  From (\ref{spectrum}), we notice a massless mode with $m^2=0$, which agrees with the results of the last subsection. There is an easier way to see that there is a massless mode. Clearly, $H(r)=1$ and $m^2=0$ are solutions to EOM (\ref{EOMMBCmassiveH}) and BCs (\ref{NBCH1},\ref{NBCH2}). Furthermore, this massless mode is normalizable
 \begin{eqnarray}\label{sect2:normalizable}
\int_{-\rho_1}^{\rho_2} dr \cosh^{d-2}(r) H(r)^2\ \text{is finite}. 
\end{eqnarray}
Thus, there is indeed a physical massless gravity on the brane in wedge holography with DGP terms. On the other hand, the massless mode is non-normalizable due to the infinite volume in the usual double holography 
 \begin{eqnarray}\label{sect2:non-normalizable}
\int_{-\infty}^{\rho_2} dr \cosh^{d-2}(r) H(r)^2 \to \infty. 
\end{eqnarray}

Naively, one can check that $m^2=-(d-2)$ is also a solution to (\ref{spectrum}). However, this is not the case. According to \cite{Chu:2021mvq}, (\ref{massiveHsolution}) is no-longer the general solution for $m^2=-(d-2)$. Instead, one should re-solve EOM (\ref{EOMMBCmassiveH}) with $m^2=-(d-2)$ to get the general solution. One can check this solution does not satisfy the NBCs (\ref{NBCH1},\ref{NBCH2}) at fixed brane positions. Instead, they correspond to the brane bending modes, allowing the brane positions to change. We will discuss the brane bending modes in the next subsection. To end this subsection, we list the mass spectrum in Table. \ref{table1spectrum} and Table. \ref{table2spectrum} below. Without loss of generality, we take $\rho_a=0.5, \lambda_a=0.1$ and $\rho_a=0.5, \lambda_a=-0.1$ as examples. Table. \ref{table1spectrum} and Table. \ref{table2spectrum} show that the mass $m$ and mass gap $\Delta m$ become larger for negative $\lambda_a$. Thus, Einstein's gravity is a better approximation at the low energy scale as the brane effective theory for negative $\lambda_a$. That is because the massive mode is more difficult to be excited due to the more significant mass gap for negative $\lambda_a$. 

\begin{table}[ht]
\caption{Mass spectrum for $d=3$}
\begin{center}
    \begin{tabular}{| c | c | c | c |  c | c | c | c| c| c|c| }
    \hline
     & $1$ & $2$ & 3  & 4& 5 \\ \hline
  $m^2$ for $\lambda_a=0.1$   & 0 & 5.124 & 25.011 & 61.667& 117.415 \\ \hline
  $m^2$ for $\lambda_a=-0.1$  & 0 & 22.511 & 74.747 & 149.216& 245.281  \\ \hline
    \end{tabular}
\end{center}
\label{table1spectrum}
\end{table}

\begin{table}[ht]
\caption{Mass spectrum for $d=4$}
\begin{center}
    \begin{tabular}{| c | c | c | c |  c | c | c | c| c| c|c| }
    \hline
     & $1$ & $2$ & 3  & 4& 5 \\ \hline
  $m^2$ for $\lambda_a=0.1$   & 0 & 4.776 & 24.950 & 61.837& 117.742 \\ \hline
  $m^2$ for $\lambda_a=-0.1$  & 0 & 22.721 & 75.211 & 149.764& 245.866  \\ \hline
    \end{tabular}
\end{center}
\label{table2spectrum}
\end{table}

\subsection{Brane bending mode}

Let's study the brane bending modes \cite{Garriga:1999yh, Izumi:2022opi}. In the last subsection, we focus on the fixed brane locations (\ref{perturbationQ1}). In general, there are fluctuations for the brane positions
 \begin{eqnarray}
Q_1: \ r=-\rho_1-\epsilon\ \phi_1(y), \ \ \ \ \ Q_2: r=\rho_2-\epsilon\ \phi_2(y). \label{bendingmodeQ}
\end{eqnarray}
We assume that the metric perturbation is still given by (\ref{perturbationmetric}) with the gauge (\ref{hij1gauge}).  By using  (\ref{perturbationmetric}, \ref{hij1gauge}, \ref{EOMMBCmassivehij},\ref{bendingmodeQ}), we can simplify the NBC (\ref{NBC}) as
 \begin{eqnarray}
 &&\Big(-\frac{1}{2} \cosh^2(\rho_1) H'(-\rho_1) -\lambda_1 H(-\rho_1)
m^2\Big) h^{(1)}_{ij}\nonumber\\
&&-\Big( D_i D_j\phi_1-(\Box+(1-d))\phi_1 h^{(0)}_{ij}\Big) (1+2(d-2)\lambda_1 \tanh\rho_1)=0, \label{NBCQ1}\\
&& \Big( \frac{1}{2} \cosh^2(\rho_2) H'(\rho_2) -\lambda_2 H(\rho_2)
m^2\Big) h^{(1)}_{ij}\nonumber\\
&&+\Big( D_i D_j\phi_2-(\Box+(1-d))\phi_2 h^{(0)}_{ij}\Big) (1+2(d-2)\lambda_2 \tanh\rho_2)=0, \label{NBCQ2}
 \end{eqnarray}
where $\Box$ is the d'Alembert operator defined by $h^{(0)}_{ij}$.  Note that (\ref{NBCQ1},\ref{NBCQ2}) agree with (\ref{NBCH1},\ref{NBCH2}) at fixed brane locations, i.e., $\phi_1=\phi_2=0$. Taking the trace of (\ref{NBCQ1}, \ref{NBCQ2}) and using $h^{(1)i}_{\ \ \ \ i}=0$, we derive
  \begin{eqnarray}\label{phi2}
(\Box -d )\phi_a=0,
 \end{eqnarray}
 where $a$ denotes $1,2$. 
The traceless parts of (\ref{NBCQ1}, \ref{NBCQ2}) give
  \begin{eqnarray}\label{h1ijQ1}
&&h^{(1)}_{ij}= \frac{-2 \left(2 (d-2) \lambda _1 \tanh \left(\rho _1\right)+1\right)}{\cosh ^2\left(\rho _1\right) H'\left(-\rho _1\right)+2 \lambda_1 m^2 H\left(-\rho _1\right)}\  \Big(D_i D_j -\frac{1}{d}h^{(0)}_{ij}\Box\Big) \phi_1,\\
&&h^{(1)}_{ij}= \frac{2 \left(2 (d-2) \lambda _2 \tanh \left(\rho _2\right)+1\right)}{2 \lambda _2 m^2 H\left(\rho _2\right)-\cosh ^2\left(\rho _2\right) H'\left(\rho _2\right)} \  \Big(D_i D_j -\frac{1}{d}h^{(0)}_{ij}\Box\Big) \phi_2 \label{h1ijQ2},
 \end{eqnarray}
 which implies that $\phi_1$ and $\phi_2$ are not independent generally.  Substituting either (\ref{h1ijQ1}) or (\ref{h1ijQ2}) into (\ref{EOMMBCmassivehij}), we derive
   \begin{eqnarray}\label{mmbendingmode}
m^2=-(d-2),
 \end{eqnarray}
where we have the following formula  \cite{Izumi:2022opi} in the above calculations
  \begin{eqnarray}\label{bendingmodeformula}
\Big(\Box+2+(d-2) \Big) \Big(D_i D_j -\frac{1}{d}h^{(0)}_{ij}\Box \Big)\phi_a=\Big(D_i D_j -\frac{1}{d}h^{(0)}_{ij}\Box\Big)(\Box-d)\phi_a=0.
 \end{eqnarray}
Thus, the brane bending modes produce a metric perturbation (\ref{h1ijQ1},\ref{h1ijQ2}) with $m^2=-(d-2)$.

Note that, for the ansatz of bulk metric (\ref{perturbationmetric}) with gauge (\ref{hij1gauge}), the bending modes $\phi_1$ and $\phi_2$ are not independent. One may consider a more general ansatz of the metric perturbation with non-zero $\delta g_{r i}$ to have independent brane bending modes \footnote{Near one brane, we may remove $\delta g_{r i}$ by suitable coordinate transformations. However, generally, one cannot delete $\delta g_{r i}$ near both branes. }. Or equivalently, one chooses two coordinate patches, the first (second) of which includes only the left (right) brane. In each coordinate patch, the bulk metric is still given by (\ref{perturbationmetric}). An additional coordinate transformation is needed to relate the metrics in the overlap of these patches. See \cite{Charmousis:1999rg} for more discussions. For simplicity, we focus on the case of one independent brane bending mode in this paper. We discuss the left and right bending modes, respectively. Take the left one as an example. We choose
  \begin{eqnarray}
Q_1: \ r=-\rho_1-\epsilon\  \phi_1(y), \ \ \ \ \ Q_2: r=\rho_2, \label{bendingmodeQ1}
\end{eqnarray}
and impose the BC for $H(r)$
  \begin{eqnarray} \label{bendNBCHQ1}
 && \frac{-2 \left(2 (d-2) \lambda _1 \tanh \left(\rho _1\right)+1\right)}{\cosh ^2\left(\rho _1\right) H'\left(-\rho _1\right)+2 \lambda_1 m^2 H\left(-\rho _1\right)}=1,\\ 
&&\Big( \frac{1}{2} \cosh^2(\rho_2) H'(\rho_2) -\lambda_2 H(\rho_2) \label{bendNBCHQ2}
m^2\Big)=0.
\end{eqnarray}
Then we get the metric perturbation (\ref{h1ijQ1})
  \begin{eqnarray}\label{lefth1ijQ1}
h^{(1)}_{ij}= \Big(D_i D_j -\frac{1}{d}h^{(0)}_{ij}\Box\Big) \phi_1
 \end{eqnarray}
with $m^2=-(d-2)$.  Similarly, one can obtain the bending mode for the right brane. 

Now let us study the effective action for the brane bending mode. It is more convenient to take another ansatz of the bulk metric instead of (\ref{perturbationmetric}). By performing suitable coordinate transformations, one can rewrite the metric (\ref{perturbationmetric}) with $h^{(1)}_{ij} \sim  \Big(D_i D_j -\frac{1}{d}h^{(0)}_{ij}\Box\Big) \phi$ into the following form \cite{Charmousis:1999rg}. See also \cite{Kanno:2002ia}.
\begin{eqnarray}\label{newbendingmetric}
ds^2=\Big(1+\epsilon H_1(r)\phi(y) \Big) dr^2+ \Big(1+\epsilon H_2(r)\phi(y) \Big) \cosh^2(r) h^{(0)}_{ij} dy^i dy^j,
 \end{eqnarray}
where $\phi(y)$ denotes the brane bending mode, $H_1(r)$ and $H_2(r)$ are functions to be determined.  Compared with (\ref{perturbationmetric}), the metric (\ref{newbendingmetric}) has the advantage that it includes less derivatives of $\phi$. Solving Einstein equations at the linear order, we obtain 
\begin{eqnarray}\label{newbendingH1H2}
H_1(r)=-c_1(d-2) \text{sech}^{d-2}(r), \  H_2(r)=c_1 \text{sech}^{d-2}(r),
 \end{eqnarray}
 and 
   \begin{eqnarray}\label{newbendingphi}
(\Box -d )\phi=0,
 \end{eqnarray}
 where $c_1$ is an integral constant. Comparing (\ref{newbendingphi}) with (\ref{phi2}), we see that $\phi$ obeys the EOM of the brane bending mode. In fact, $\phi$ indicates the relative motion of the two branes, which is called the radion  \cite{Charmousis:1999rg,Kanno:2002ia}. 
 
Let us go on to derive the location of the two branes. Substituting the embedding functions (\ref{bendingmodeQ}) into the NBC (\ref{NBC}), we solve
   \begin{eqnarray}\label{newbendingphi1phi2}
\phi_1=-\frac{c_1 (d-2) \lambda _1 \text{sech}^{d-2}\left(\rho _1\right)}{1+2 (d-2) \lambda _1 \tanh \left(\rho _1\right)} \phi , \ \ \phi_2= \frac{c_1 (d-2) \lambda _2 \text{sech}^{d-2}\left(\rho _2\right)}{1+2 (d-2) \lambda _2 \tanh \left(\rho _2\right)} \phi. 
 \end{eqnarray}
 Substituting the bulk metric (\ref{newbendingmetric},\ref{newbendingH1H2}) together with the embedding functions of branes (\ref{bendingmodeQ},\ref{newbendingphi1phi2}) into the action (\ref{action}) and integrating along $r$, we finally obtain the squared action of the radion
    \begin{eqnarray}\label{action of radion}
I_{\phi}= B \ c_1^2 \epsilon^2 \int dy^d\sqrt{-h^{(0)}} \Big(-\frac{1}{2}D_i \phi D^i \phi- \frac{d}{2} \phi^2 \Big) ,
 \end{eqnarray}
 where $B$ is given by
    \begin{eqnarray}\label{bendmodeB}
B=\sum_{a=1}^2\Big(\ \frac{(d-1) (d-2) }{2} \int_0^{\rho_a} dr \ \text{sech}^{d-2}(r)+ \frac{(d-1) (d-2) \text{sech}^{d-2}\left(\rho _a\right)}{1+2 (d-2) \lambda _a \tanh \left(\rho _a\right)}\lambda _a \  \Big).
 \end{eqnarray}
 Note that we have drop some total derivative terms in the above derivations. In particular, the linear action of $\phi$ is a total derivative as expected.  From the action (\ref{action of radion}), we can derive the correct EOM of $\phi$ (\ref{newbendingphi}). This can be regarded as a test of our calculations. To have the positive kinetic energy, we require that
 \begin{eqnarray}\label{condition of kinetic energy}
B \ge 0,
 \end{eqnarray}
which imposes another constraint on the parameters $(\rho_a,\lambda_a)$.  Now we have obtained two constraints (\ref{boundDGP1}) and (\ref{condition of kinetic energy}) for the parameters of our model.


\subsection{Holographic entanglement entropy}

\begin{figure}[t]
\centering
\includegraphics[width=12cm]{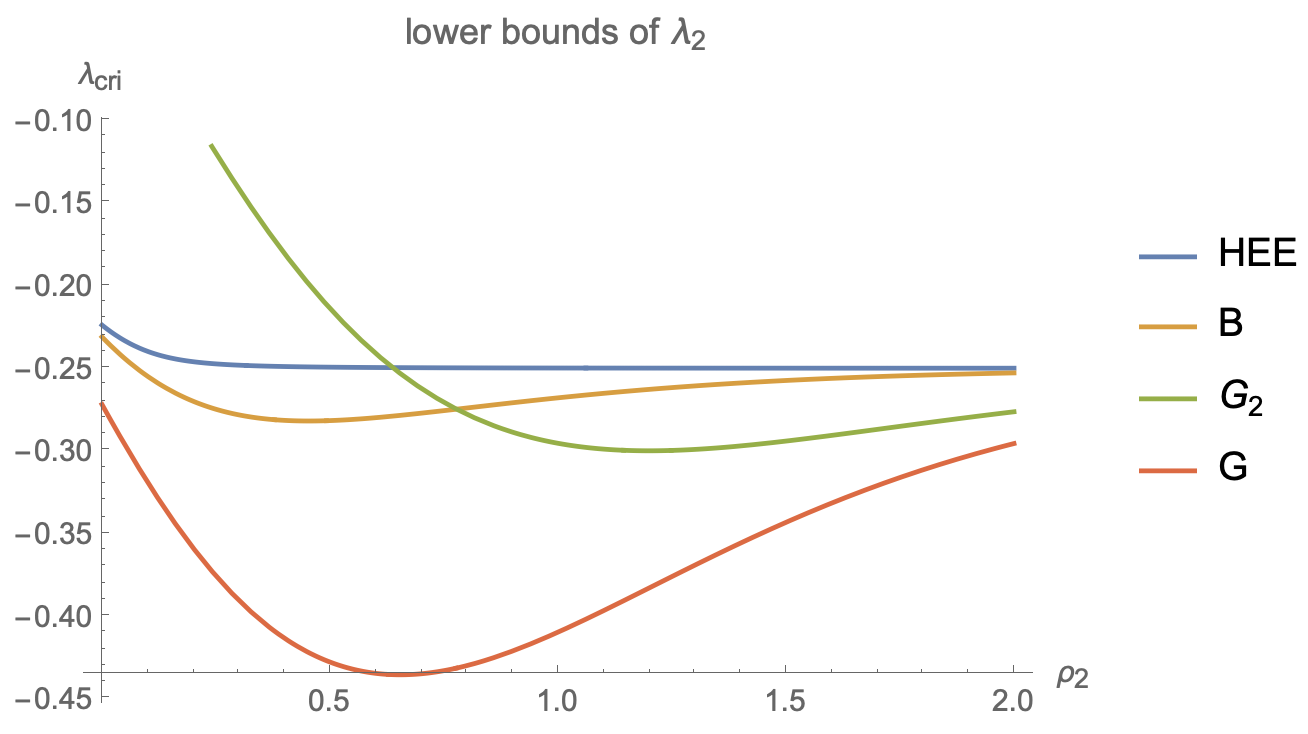}
\caption{Various lower bounds of the DGP parameter $\lambda_2 $ for $\rho_1=0.5, \lambda_1=0, d=4$, i.e., $\lambda_2\ge \lambda_{\text{cri}}$. The blue, orange, green, red curves denote the lower bounds derived from HEE (\ref{sect2.4:lambdaHEE}), brand bending modes (\ref{bendmodeB}), effective Newton's constants (\ref{effective Newton constant}), respectively.  Here $\frac{1}{G}=\frac{1}{G_1}+\frac{1}{G_2}$, and $G_1, G_2$ are the effective Newton's constants on the two branes. It shows that the Newton's constant $G_2$ and HEE impose the strongest constraint for $\rho_2 < 0.638$, and $\rho_2 > 0.638$, respectively. In the large tension limit $\rho_2 \to \infty$, all lower bounds approach to $ \lambda_{\text{cri}}\to -1/(2(d-2))$. }
\label{lowerboundcaseI}
\end{figure}

\begin{figure}[t]
\centering
\includegraphics[width=12cm]{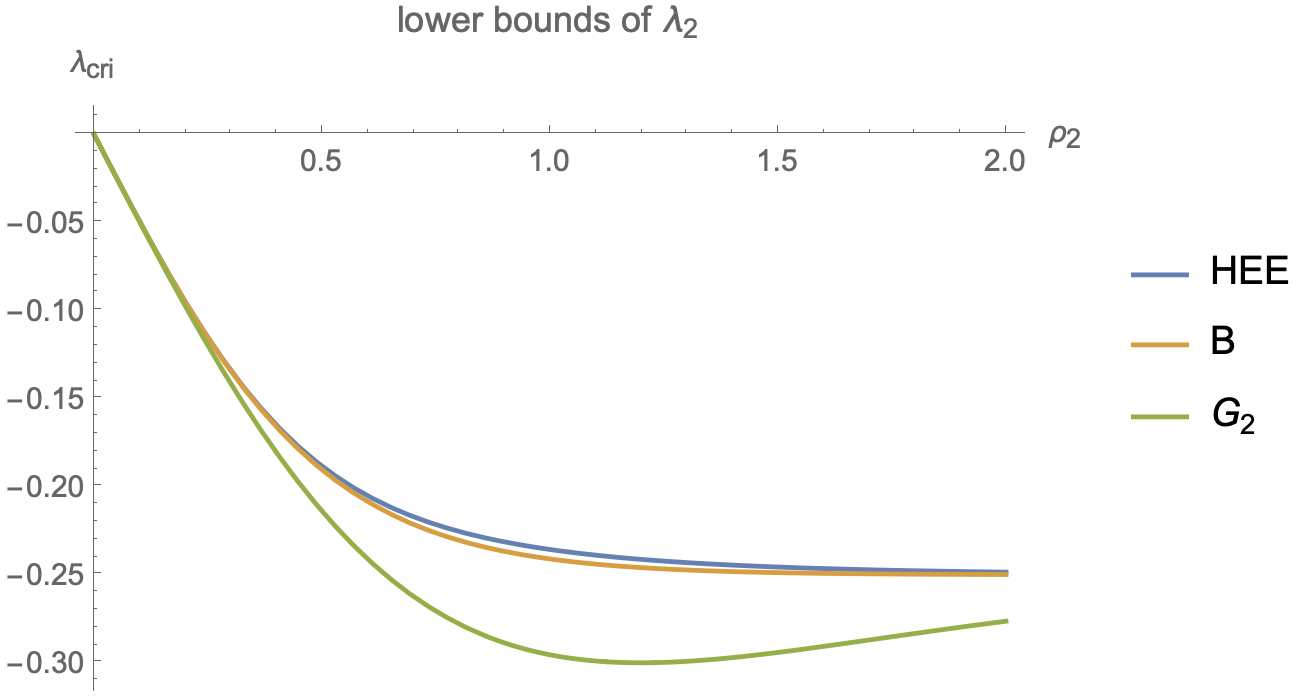}
\caption{Various lower bounds of the DGP parameter $\lambda_2 $ for $\rho_1=0, \lambda_1=0, d=4$, i.e., $\lambda_2\ge \lambda_{\text{cri}}$. The blue, orange, green curves denote the lower bounds derived from HEE (\ref{sect2.4:lambdaHEE}), brane bending modes (\ref{bendmodeB}), effective Newton's constants (\ref{effective Newton constant}), respectively.  It shows that the HEE impose the strongest lower bound of $\lambda_2$. In the large tension limit $\rho_2 \to \infty$, all lower bounds approach to $ \lambda_{\text{cri}}\to -1/(2(d-2))$. }
\label{lowerboundcaseII}
\end{figure}

In this subsection, we study the holographic entanglement entropy (HEE) \cite{Ryu:2006bv} for CFTs on the $(d-1)$-dimensional defect $\Sigma$ in wedge holography with DGP gravity. We focus on the vacuum state on the whole defect $\Sigma$ for simplicity. Since it is a pure state, the HEE is expected to be zero \footnote{ Note that we are studying regularized finite HEE since the branes locate at a finite place instead of infinity. Similar to Casimir energy, the regularized HEE can be negative in principle. As a result, we can relax the constraint and require that the HEE is bounded from below. Interestingly, this relaxed constraint yields the same lower bound for the DGP parameter as zero HEE.}, which causes another lower bound of the DGP parameter.

From the action (\ref{action}), we read off HEE
\begin{eqnarray}\label{HEE}
S_{\text{HEE}}={\text{min}} \Big\{ \text{ext} \Big( 4\pi \int_{\Gamma} dx^{d-1} \sqrt{\gamma}+8\pi \int_{\partial \Gamma} dx^{d-2} \sqrt{\sigma} \lambda_a \Big)\Big\},
\end{eqnarray}
where $\Gamma$ denote the RT surface in the bulk, $\partial \Gamma=\Gamma\cap Q$ is the intersection of the RT surface and the brane $Q$, $\gamma$ and $\sigma$ represent the induced metric on $\Gamma$ and $\partial \Gamma$ respectively. Since we are interested in the vacuum state of the defect, we focus on the AdS space (\ref{AdSmetric}) in bulk. Substituting the embedding functions $z=z(r)$ and $t=\text{constant}$ into the AdS metric (\ref{AdSmetric}) and entropy formula (\ref{HEE}), i.e., $S_{\text{HEE}}=4\pi A$, we get the area functional of RT surface
\begin{eqnarray}\label{sect2.4:area}
A=\int_{-\rho_1}^{\rho_2} dr\frac{\cosh^{d-2}(r)}{z(r)^{d-2}} \sqrt{1+\frac{\cosh^{2}(r) z'(r)^2}{z(r)^{2}}}+\sum_{a=1}^2\frac{2\lambda_a \cosh^{d-2}(\rho_a)}{z^{d-2}_a},
\end{eqnarray}
where we have set the tangential volume $V=\int dy^{d-2}=1$, and $z_a=z( (-)^a \rho_a)$ denotes the endpoints of the RT surfaces on the branes. Taking variations of (\ref{sect2.4:area}), we derive the Euler-Lagrange equation
\begin{eqnarray}\label{sect2.4:ELEOM}
&&z^2 \cosh (r) \left(d \sinh (r) z'+\cosh (r) z''\right)+(d-2) z^3\nonumber\\
&&+(d-3) z \cosh ^2(r) \left(z'\right)^2+(d-1) \sinh (r) \cosh ^3(r) \left(z'\right)^3=0
\end{eqnarray}
and NBC on the branes
\begin{eqnarray}\label{sect2.4:NBC}
\frac{(-)^az_a'}{\sqrt{z_a^2+\cosh^{2}(\rho_a) z_a'^2}}=\frac{2\lambda_a(d-2)}{\cosh^2(\rho_a)}.
\end{eqnarray}

Note that the AdS metric (\ref{AdSmetric}) is invariant under the rescale $z\to c z$. 
Due to this rescale invariance, if $z=z_0(r)$ is an extremal surface, so does $z=c z_0(r)$. Under the rescale $z\to c z$, the area functional (\ref{sect2.4:area}) transforms as $A \to A/c^{d-2}$.  
Recall that the RT surface is the extremal surface with minimal area. By choosing $c\to \infty$, we get the RT surface $z=c z_0(r)\to \infty $ with zero area $A=A_0/c^{d-2}\to 0$, provided $A_0$ is positive. Here $A_0$ denotes the area of the input extremal surface $z=z_0(r)< \infty$. On the other hand, if $A_0$ is negative for sufficiently negative $\lambda_a$, the RT surface is given by choosing $c\to 0$ so that $A=A_0/c^{d-2}\to -\infty$. To rule out this unusual case with negative infinite entropy, we must impose a lower bound on $\lambda_a$.

For simplicity, we focus on the case with $\lambda_1=0$ and discuss how to derive the lower bound of $\lambda_2$. The approach is as follows. We take an arbitrarily start point $0<z_1=z(-\rho_1)<\infty$ on the left brane $Q_1$, and impose the orthogonal condition $z'_1=z'(-\rho_1)=0$, then we solve EOM (\ref{sect2.4:ELEOM}) to determine the extremal surface $z=z_0(r)$ numerically. By requiring the corresponding area $A_0$ (\ref{sect2.4:area}) is non-negative, we obtain a lower bound
\begin{eqnarray}\label{sect4:lambdabound}
\lambda_2\ge \lambda_{\text{HEE}},
\end{eqnarray}
where $\lambda_{\text{HEE}}$ is derived from $A_0=0$. Note that $A_0=0$ means that the corresponding extremal surface is the RT surface with minimal area. As a necessary condition, it should satisfy the NBC (\ref{sect2.4:NBC}) on the right brane $Q_2$. From (\ref{sect2.4:NBC}), we derive
\begin{eqnarray}\label{sect2.4:lambdaHEE}
\lambda_{\text{HEE}}(\rho_2)=\frac{\cosh ^2(\rho_2) z'_2}{2 (d-2)\sqrt{\cosh ^2(\rho_2 ) z_2'^2+z_2^2}},
\end{eqnarray}
where $z_2=z(\rho_2)$ is the endpoint on the right brane $Q_2$. Due to the rescale invariance of AdS, any input start point $z_1=z(-\rho_1)$ gives the same $\lambda_{\text{HEE}}$ (\ref{sect2.4:lambdaHEE}). In other words, there are infinite zero-area RT surfaces, which obey NBCs on both branes. It is similar to the case of $\text{AdS}_3$ in AdS/BCFT. On the other hand, for $\lambda>\lambda_{\text{HEE}}$, the RT surface locates only at infinity, i.e., $z\to \infty$. And the NBC (\ref{sect2.4:NBC}) can be satisfied only at infinity for $\lambda_2>\lambda_{\text{HEE}}$. Please see blue curves of Fig.\ref{lowerboundcaseI} and Fig.\ref{lowerboundcaseII} for the lower bound $\lambda_{\text{HEE}}(\rho_2)$. In Fig.\ref{lowerboundcaseI} with $\rho_1=0.5$ and $d=4$, we notice that Newton's constant $G_2$ imposes the strongest lower bound for $\rho_2 < 0.638$, while HEE imposes the strongest lower bound for $\rho_2 > 0.638$. In Fig. \ref{lowerboundcaseII} with $\rho_1=0$ and $d=4$, we find that HEE always gives the strongest lower bound for $\lambda_2$. 

In summary, we have discussed various constraints of the DGP parameters from effective Newton's constants, brane bending modes, and HEE. We find that HEE imposes the strongest lower bound of $\lambda_2$ for sufficiently large $\rho_2$.

\section{Page curve of case I: one black hole approximately}

\begin{figure}[t]
\centering
\includegraphics[width=9cm]{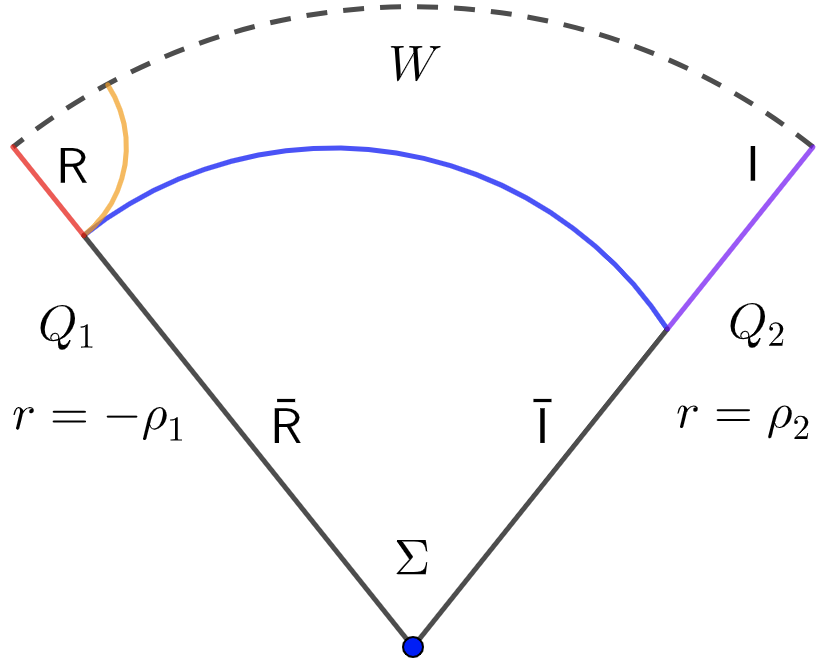}
\caption{Geometry for case I: one black hole approximately. $Q_1$ denotes the bath brane with weak gravity, and $Q_2$ is the AdS brane with intense gravity. The red and black lines denotes the radiation $\text{R}$ and its complement $\bar{\text{R}}$ on the left brane, the purple and black lines denotes the island $\text{I}$ and its complement $\bar{\text{I}}$ on the right brane. The island region $\text{I}$ and radiation region $\text{R}$ envelop the black-hole horizon on $Q_1\cup Q_2$. For simplicity, we only show the regions outside the horizon. The dotted line, blue, and orange lines in the bulk indicate the horizon, RT surface in the island phase and HM surface in the no-island phase at $t=0$, respectively. }
\label{Wedge1}
\end{figure}

\begin{figure}[t]
\centering
\includegraphics[width=10cm]{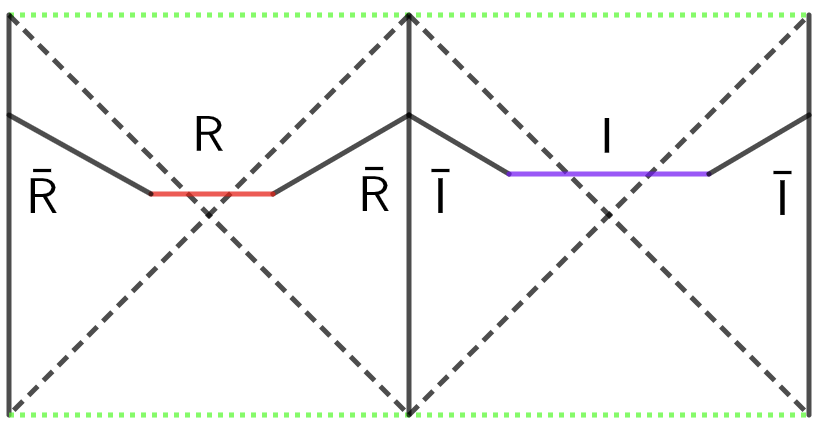}
\caption{Penrose diagram on the branes for case I: one two-side black hole approximately. The left and right vertical lines are glued together.  The black-dotted, green-dotted, red, and purple lines denote the horizon, singularity, radiation region, and island region, respectively. The black lines linking R and I represent $\bar{\text{R}}\cup \bar{\text{I}}$.}
\label{Penrose1}
\end{figure}

The above section investigates some aspects of wedge holography with DGP gravity (DGP wedge holography) on the branes. In particular, we find that there is massless gravity on the branes, and we get 
several constraints (\ref{boundDGP1},\ref{condition of kinetic energy},\ref{sect4:lambdabound}) 
for the parameters $(\rho_a,\lambda_a)$. This section studies the Page curve in DGP wedge holography for case I.  We focus on the eternal two-side black hole, which is dual to the thermofield double state of CFTs \cite{Maldacena:2001kr}. See Fig. \ref{Wedge1} for one side of the system at the time slice $t=0$. See also Fig. \ref{Penrose1} for the Penrose diagram of the two-side black holes on the branes. 

Let us focus on the black string in bulk
\begin{eqnarray}\label{sect3:BHmetric}
ds^2=dr^2+\cosh^2(r) \frac{\frac{dz^2}{f(z)}-f(z)dt^2+\sum_{\hat{i}=1}^{d-2}dy_{\hat{i}}^2}{z^2},\ \ -\rho_1\le r \le \rho_2,
\end{eqnarray}
where $f(z)=1-z^{d-1}$, the weak-gravity brane $Q_1$ and strong-gravity brane $Q_2$ locate at $r=-\rho_1$ and $r=\rho_2$, respectively. See Fig. \ref{Wedge1} for the geometry. Note that there are two black holes on the branes $Q_1\cup Q_2$. 
Following \cite{Geng:2020fxl}, we take the black hole on the weak-gravity brane $Q_1$ as the bath approximately. Since both branes are gravitating, we should adjust both the radiation region $\text{R}$ (red line) and the island region $\text{I}$ (purple line) to minimize the entanglement entropy of the radiation $\text{R}$ \cite{Geng:2020fxl}. Once this approach determines the radiation region $\text{R}$, we can follow the usual procedure to calculate the entanglement entropy of $\text{R}$, which is given by the Hartman-Maldacena (HM) surface (orange curve of Fig.\ref{Wedge1}) at early times and given by RT surface in the island phase (blue curve of Fig.\ref{Wedge1}) at late times.

To warm up, we start with wedge holography without DGP gravity. See Fig. \ref{wedge without DGP} for the geometry. We first investigate the island phase, where the RT surface ends on the branes and stays outside the horizon. Assuming the embedding function $z=z(r)$ and $t=\text{constant}$, from (\ref{sect3:BHmetric}) we derive the area functional of the RT surface 
\begin{eqnarray}\label{sect3:areaislandnoDGP}
A_{\text{I}}=V\int_{-\rho_1}^{\rho_2} dr\frac{\cosh^{d-2}(r)}{z(r)^{d-2}} \sqrt{1+\frac{\cosh^{2}(r) z'(r)^2}{z(r)^{2}f(z(r))}},\end{eqnarray}
where $\text{I}$ means the island phase, and $V=\int dy^{d-2}$ denotes the tangential volume.  From $0\le z(r) \le 1$ and $f(z)\ge 0$, we derive an inequality
\begin{eqnarray}\label{sect3:areaislandnoDGP1}
A_{\text{I}}\ge V\int_{-\rho_1}^{\rho_2} dr \cosh^{d-2}(r) = A_{\text{BH}},
\end{eqnarray}
where $A_{\text{BH}}$ is the horizon area in bulk. The above inequality shows that the area functional (\ref{sect3:areaislandnoDGP}) minimizes on the horizon $z(r)=1$. In other words, as shown in Fig.\ref{wedge without DGP}, the RT surface (minimal area surface) coincides with the horizon  in the island phase. Comparing Fig. \ref{doubleholography} with Fig. \ref{wedge without DGP}, we notice that the island region $\text{I}$ in the usual double holography disappears in wedge holography without DGP terms \footnote{We mean the region outside the horizon disappears.}.  Now we reproduce the results of \cite{Geng:2020fxl} in a simpler method.

\subsection{Island phase}

Let us go on to discuss the island phase in wedge holography with DGP gravity on the branes. We find a non-trivial RT surface outside the horizon and, thus a non-trivial island region for suitable DGP terms. See Fig. \ref{Wedge1} for the geometry at a time slice. 

For the DGP wedge holography, the area functional becomes
\begin{eqnarray}\label{sect3.1:areaisland}
A_{\text{I}}=\frac{S_{\text{HEE}}}{4\pi}=V\int_{-\rho_1}^{\rho_2} dr\frac{\cosh^{d-2}(r)}{z(r)^{d-2}} \sqrt{1+\frac{\cosh^{2}(r) z'(r)^2}{z(r)^{2}f(z(r))}}+V\sum_{a=1}^2\frac{2\lambda_a \cosh^{d-2}(\rho_a)}{z^{d-2}_a},
\end{eqnarray}
where $z_a=z( (-)^a \rho_a)$ denotes the endpoints of the RT surfaces on the branes. To have a well-defined variational principle for (\ref{sect3.1:areaisland}), we can impose either Dirichlet boundary condition (DBC) $\delta z_a=0$ or NBC on each brane
\begin{eqnarray}\label{sect3.1:NBCisland}
\frac{(-)^az_a'}{f(z_a)\sqrt{1+\frac{\cosh^{2}(\rho_a) z_a'^2}{z_a^{2}f(z_a)}}}=\frac{2\lambda_a(d-2)z_a}{\cosh^2(\rho_a)}.
\end{eqnarray}
Usually, NBC yields a smaller area than DBC since it allows the endpoint of the RT surface to move on the brane. From (\ref{sect3.1:areaisland}), we derive the Euler-Lagrangian equation 
 \begin{eqnarray}\label{sect3.1:ELEOM}
&&\cosh ^2(r) \left(z'\right)^2 \left((d-1) \sinh (2 r) z'-(d-5) z^d+2 (d-3) z\right)\nonumber\\
&&+2 (d-2) z \left(z-z^d\right)^2-2 z \left(z^d-z\right) \cosh (r) \left(d \sinh (r) z'+\cosh (r) z''\right)=0,
\end{eqnarray}
where we abbreviate $z(r)$ by $z$ to simplify the above equation. 

Note that the first term of the area functional (\ref{sect3.1:areaisland}) decreases with $z(r)$, while the second term of (\ref{sect3.1:areaisland}) increases with $z_a$ for negative DGP parameters $\lambda_a$. These two terms compete and can yield non-trivial RT surfaces outside the horizon, i.e., $z(r)<1$.  As a result, the island region becomes non-zero for negative $\lambda_a$, as shown in Fig. \ref{Wedge1}.   

Let us take the method of \cite{Geng:2020fxl} to understand why DGP gravity can recover the entanglement islands. Without the DGP terms, as a minimal area surface, the RT surface should end orthogonally on both branes, i.e., $z'_a=0$. This orthogonal condition rules out all of the extremal surfaces except the horizon \cite{Geng:2020fxl}. When the DGP gravity appears, the orthogonal condition breakdowns \cite{Chen:2020uac}, i.e., $z'_a\ne 0$. See the NBC (\ref{sect3.1:NBCisland}), which gives $z'_a\sim \lambda_a\ne 0$. As a result, the no-go theorem based on $z'_a=0$ disappears. That is why there could be non-trivial RT surfaces outside the horizon, equivalently, non-vanishing entanglement islands.  

Now we show how to construct the RT surface outside the horizon exactly. To do so, we turn the logic around. Suppose we have solved a series of extremal surfaces outside the horizon from the EOM (\ref{sect3.1:ELEOM}). For any extremal surface, we can derive $z_a, z_a'$ on the branes and obtain $\lambda_a$ from NBC (\ref{sect3.1:NBCisland}). Let us return to our problem. For the DGP parameters $\lambda_a$ fixed above, the RT surface is just the input extremal surface outside the horizon because it satisfies both the Euler-Langrangian equation (\ref{sect3.1:ELEOM}) and NBC (\ref{sect3.1:NBCisland}). We should further check that the extremal surface is minimal instead of maximal. As shown below, we can always do so by choosing suitable parameters. Now we finish the construction of entanglement islands in wedge holography with DGP gravity on the branes.

Let us study an exact example. Without losing generality, we choose the parameters
\begin{eqnarray}\label{sect3.1:parameters}
\rho_1=0.5, \lambda_1=0,\rho_2=1.2, \lambda_2\approx -0.246,\ d=4,\  V=1,
\end{eqnarray}
which obey the constraints from Newton's constants (\ref{boundDGP1})
\begin{eqnarray}\label{sect3.1:G1G2}
0<G^1_{\text{eff N}} \approx 0.037 < G^2_{\text{eff N}}\approx 0.056,
\end{eqnarray}
brane bending modes (\ref{condition of kinetic energy}) and HEE (\ref{sect4:lambdabound})
\begin{eqnarray}\label{c2}
B \approx 1.391 >0, \ \ \lambda_2 > \lambda_{\text{HEE}}\approx -0.250.
\end{eqnarray}
Note that we show only three valid digits after the decimal point in this paper. In the numerical calculations, we keep more valid numbers. For instance, we have $\lambda_2\approx -0.245829$. 

\begin{figure}[t]
\centering
\includegraphics[width=9cm]{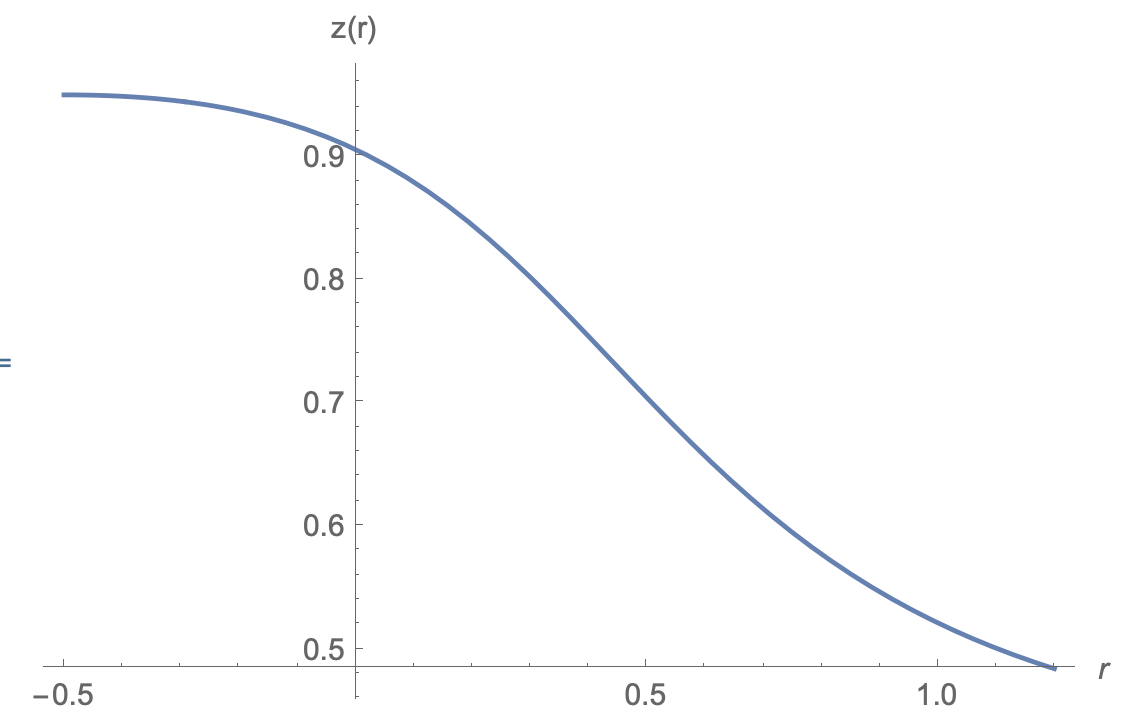}
\caption{The RT surface $z(r)$ in the island phase, which starts at $z_1\approx 0.950$ on the left brane $r=-0.5$ and ends at $z_2\approx 0.484$ on the right brane $r=1.2$. It shows that $z'_1\sim -\lambda_1=0$ and  $z'_2\sim \lambda_2<0$, which agrees with our parameters (\ref{sect3.1:parameters}) and NBC (\ref{sect3.1:NBCisland}). }
\label{zr}
\end{figure}

\begin{figure}[t]
\centering
\includegraphics[width=7.5cm]{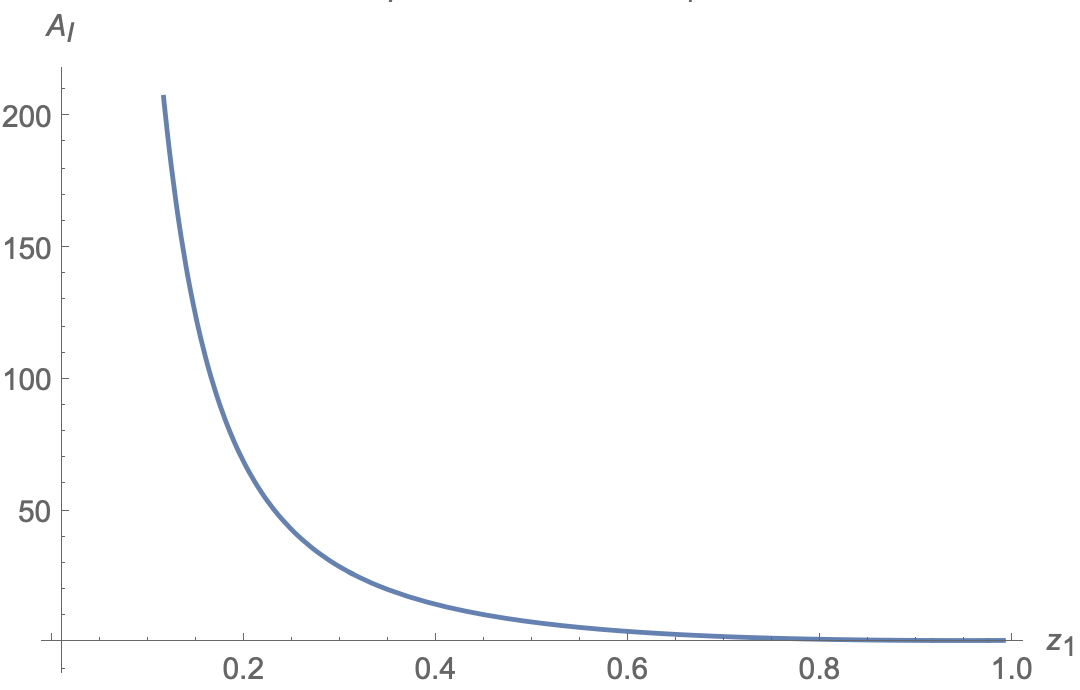}
\includegraphics[width=7.5cm]{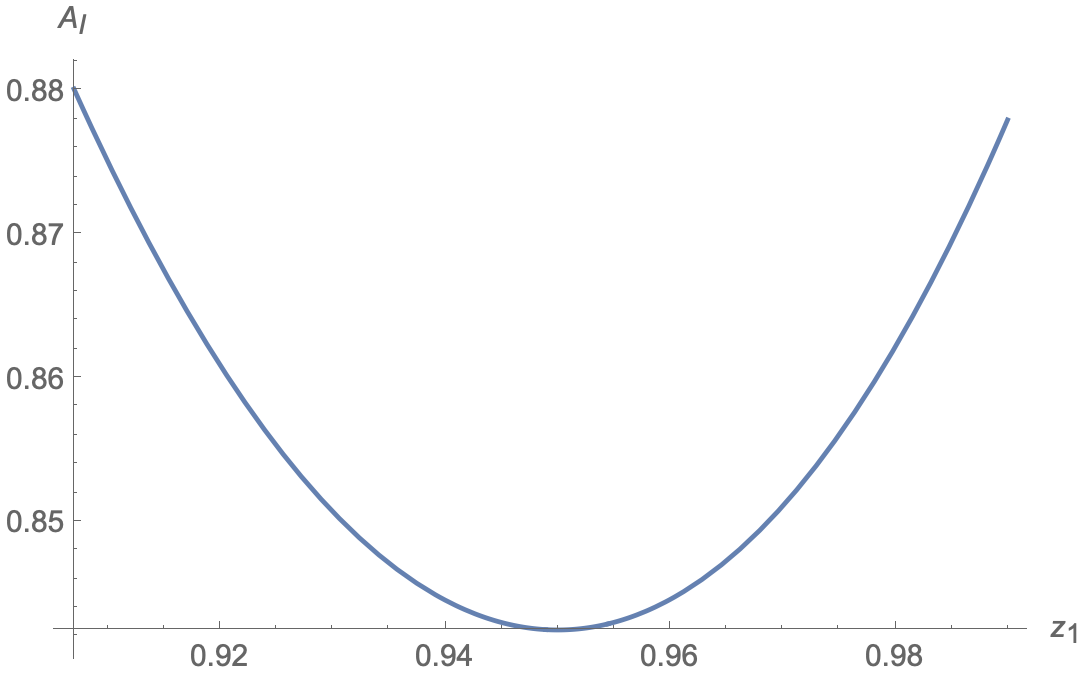}
\caption{Relation between the area $A_{\text{I}}$ and the endpoint $z_1$ on the left brane, which shows that area functional (\ref{sect3.1:areaisland}) becomes minimal at $z_1\approx 0.950$.}
\label{AIz1}
\end{figure}

Naturally, we choose the left brane as the bath brane since it has a smaller effective Newton's constant \ref{sect3.1:G1G2}, i.e., $G^1_{\text{eff N}} < G^2_{\text{eff N}}$.  Solving the Euler-Langrangian equation (\ref{sect3.1:ELEOM}) together with NBC (\ref{sect3.1:NBCisland}), we obtain numerically the RT surface $z(r)$, which starts at $z_1\approx 0.950$ on the left brane and ends at $z_2\approx 0.484$ on the right brane. Please see Fig. \ref{zr}.  Let us show more details of the numerical calculations. We impose BCs $z=z_1$ and $z_1'=0$ \footnote{ Recall that we have chosen $\lambda_1=0$, which gives $z_1'=0$ from (\ref{sect3.1:NBCisland}).} on the left brane, and adjust the left endpoint $z_1$ so that the NBC (\ref{sect3.1:NBCisland}) on the right brane is satisfied. It is the so-called shooting method. In this way, we derive the RT surface shown in Fig.\ref{zr}. There is another method to calculate the RT surface. For any given $0\le z_1\le 1$ and $z'_1=0$, we can solve the extremal surface and its area $A_{\text{I}}$ (\ref{sect3.1:areaisland}). We adjust the left endpoint $z_1$ to minimize the area $A_{\text{I}}$. See Fig. \ref{AIz1} for $A_{\text{I}}(z_1)$, which shows the area $A_{\text{I}}$ becomes minimal at $z_1\approx 0.950$. From the RT surface with $z_1\approx 0.950$, we derive the right endpoint $z_2\approx 0.484$ and its derivative $z'_2\approx -0.150$. We verify that the obtained $z_2$ and $z'_2$ satisfy the NBC (\ref{sect3.1:NBCisland}), which agrees with the first method. The second method has the advantage that it is clear that the obtained RT surface is minimal rather than maximal. See Fig. \ref{AIz1} again. 

With the above numerical results, we derive the area of RT surface
\begin{eqnarray}\label{sect3.1:AIABH}
A_{\text{I}}\approx 0.842 < A_{\text{BH}} \approx 0.898,
\end{eqnarray}
which is smaller than the black hole area. Thus there is indeed a nontrivial RT surface outside the horizon. Note that $A_{\text{BH}}$ includes the contributions from DGP terms. Note also that we focus on half of the two-side black hole in (\ref{sect3.1:AIABH}). Recall that the RT surface ends on $z_1\approx 0.950$ and $z_2\approx 0.484$ on the left and right brane, respectively. According to Fig.\ref{Wedge1}, it means the radiation region lies in $z\ge z_1\approx 0.950$ on the left brane, and the island region locates at $z\ge z_2 \approx 0.484$ on the right brane.  Clearly, the island region is non-zero in wedge holography with DGP terms.

\subsection{No-Island phase}

In this subsection, we discuss the RT surface in the no-island phase, which is also called the Hartman-Maldacena (HM) surface. The HM surface starts at $z_1\approx 0.950$ on the left weak-gravity brane, ends on the horizon at the beginning time $t=0$, and then passes the horizon at $t>0$. Let us first study the case at $t=0$ (orange line of Fig.\ref{Wedge1}). By varying the endpoint on the horizon, we get the HM surface with the minimal area
\begin{eqnarray}\label{sect3.2:Anoisland}
A_{\text{N}}\approx 0.384< A_{\text{I}}\approx 0.842, \ \text{at t=0},
\end{eqnarray}
where $\text{N}$ labels the no-island phase. Since $A_{\text{N}}<A_{\text{I}}$ at $t=0$, the no-island phase dominates at the beginning. 

As the black hole evolves, the HM surface crosses the horizon. To avoid coordinate singularities, we choose the infalling Eddington-Finkelstein coordinate $dv=dt-\frac{dz}{f(z)}$. Substituting the embedding functions $v=v(z), r=r(z)$ into the metric (\ref{sect3:BHmetric}) and entropy formula (\ref{HEE}), we get the area functional
\begin{eqnarray}\label{sect3.2:areanoisland}
A_{\text{N}}=\frac{S_{\text{HEE}}}{4\pi}=V\int_{z_1}^{z_{\text{max}}} dz\frac{\cosh^{d-2}(r)}{z^{d-2}} \sqrt{r'^2-\frac{\cosh^{2}(r)}{z^{2}}v'(2+f(z)v') },
\end{eqnarray}
and the time on the left bath brane
\begin{eqnarray}\label{sect3.2:timenoisland}
t_1=t(z_1)=-\int_{z_1}^{z_{\text{max}}} \Big( v'+\frac{1}{f(z)}\Big) dz,
\end{eqnarray}
where $r=r(z), v=v(z)$ are abbreviations, $z_{\text{max}}\ge 1$ denotes the turning point of the two-side black hole. According to \cite{Carmi:2017jqz}, we have $v'(z_{\text{max}})=-\infty$ and $t(z_{\text{max}})=0$, and $z_{\text{max}}=1$ corresponds to the beginning time $t_1=0$. For simplicity, we label $t_1$ by $t$ in this paper. Note that $A_{\text{N}}$ (\ref{sect3.2:areanoisland}) is independent of the DGP parameters $\lambda_a$. That is because we have chosen $\lambda_1=0$ in our model (\ref{sect3.1:parameters}) on the left brane. Besides, the HM surface does not intersect the right brane. As a result, no terms depend on $\lambda_2$ in the area functional (\ref{sect3.2:areanoisland}) either. From (\ref{sect3.2:areanoisland}) and $-v'(2+f(z)v')\ge 0$ \cite{Carmi:2017jqz}, we can derive an inequality
\begin{eqnarray}\label{sect3.2:ANsmall}
A_{\text{N}}\ge V\int_{z_1}^{z_{\text{max}}} dz\frac{1}{z^{d-1}} \sqrt{-v'(2+f(z)v') },
\end{eqnarray}
where the RHS is obtained by setting $r(z)=0$. We remark that $r(z)=0$ is an exact solution to the Euler-Lagrangian equations derived from (\ref{sect3.2:areanoisland}) \cite{Geng:2020fxl}. However, this solution does not obey the boundary condition on the left since the left brane is located at $r=-\rho_1$ rather than $r=0$. Instead of an exact solution, $r(z)=0$ is actually an asymptotic solution at $t\to \infty$. We observe that $r_0=r(z_{\text{max}})$ approaches zero in the large time limit. See Fig. \ref{r0time} of appendix A. Note also that, in the large time limit, the integrations around the turning point $z=z_{\text{max}}$ contribute most to the area $A_N$ (\ref{sect3.2:areanoisland}) and the time (\ref{sect3.2:timenoisland}). Thus we have 
 \begin{eqnarray}\label{sect3.2:areanoisland1}
\lim_{t\to \infty} A_{\text{N}} = \lim_{r\to 0} A_{\text{N}} = V\int_{z_1}^{z_{\text{max}}}\frac{dz}{z^{d-1}} \sqrt{-v'(2+f(z)v') }.
\end{eqnarray}
Remarkably, (\ref{sect3.2:areanoisland1}) is the same as the volume conjecture of holographic complexity \cite{Susskind:2014rva, Stanford:2014jda} for a $d-$dimensional AdS-Schwarzschild black hole. Following \cite{Carmi:2017jqz,Susskind:2014rva, Stanford:2014jda}, we obtain
\begin{eqnarray}\label{sect3.2:entropylargetime}
\lim_{t\to \infty} \frac{d A_{\text{N}}}{dt}=\frac{V}{2},
\end{eqnarray}
which yields the expected result that the HM surface area increases linearly over time at late enough times. Interestingly, the late-time growth rate (\ref{sect3.2:entropylargetime}) is the same as that of holographic complexity. This seems to imply a deep relation between entanglement entropy and complexity. This issue is worth more study in the future. Note that the late-time growth rate of $A_{\text{N}}$ is universal and is independent of the choices of parameters $(\rho_a, \lambda_a)$.

Let us provide some details on how to derive (\ref{sect3.2:entropylargetime}). Since the area functional (\ref{sect3.2:areanoisland1}) does not depend on $v(z)$ exactly, we can derive a conserved quantity
 \begin{eqnarray}\label{conserved quantity}
E_{\text{N}}=-\frac{\partial L}{\partial v'}=\frac{z^{-d} \left(1+f(z) v'\right)}{\sqrt{-\frac{ v' \left(f(z) v'+2\right)}{z^2}}}
=\frac{\sqrt{-f(z_{\max })} }{z_{\max }^{d-1}},
\end{eqnarray}
where $ A_{\text{N}}=V \int dz L$, $E_{\text{N}}$ is a constant at a fixed time, and we have used $v'(z_{\text{max}})=-\infty$ to derive the last equality of (\ref{conserved quantity}). According to \cite{Carmi:2017jqz}, the conserved quantity $E_{\text{N}}$ 
approaches to an extremum in the large time limit
 \begin{eqnarray}\label{dEdzmax}
\lim_{t\to \infty} \frac{d E_{\text{N}}}{d z_{\text{max}}} =-\frac{(d-1) z_{\max }^{-d-1} \left(\bar{z}_{\max }^d-2 \bar{z}_{\max }\right)}{2 \sqrt{\bar{z}_{\max }^{d-1}-1}}=0,
\end{eqnarray}
which yields the maximal value of $z_{\max }$
 \begin{eqnarray}\label{barzmax}
\bar{z}_{\max }=\lim_{t\to \infty} z_{\max}=2^{\frac{1}{d-1}}. 
\end{eqnarray}
From (\ref{conserved quantity}), we solve
 \begin{eqnarray}\label{dvdL}
v'(z)=\frac{E_{\text{N}} z^d \left(\sqrt{E_{\text{N}}^2 z^{2 d}+z^2 f(z)}-E_{\text{N}} z^d\right)-z^2 f(z)}{f(z) \left(E_{\text{N}}^2 z^{2 d}+z^2 f(z)\right)}.
\end{eqnarray}
By using (\ref{dEdzmax}) and (\ref{dvdL}), we get
 \begin{eqnarray}\label{dvdzmax1}
&&\lim_{t\to \infty}\frac{\partial v'(z)}{\partial z_{\max}}=\lim_{t\to \infty} \frac{\partial v'(z)}{\partial E_{\text{N}}} \frac{\partial E_{\text{N}}}{\partial z_{\max}}=0,\\ \label{dvdzmax2}
&& \lim_{t\to \infty}\frac{\partial L}{\partial z_{\max}}=\lim_{t\to \infty}\frac{\partial L}{\partial v'(z)}\frac{\partial v'(z)}{\partial z_{\max}}=0.
\end{eqnarray}
Recall that $L$ is defined by $A_{\text{N}}=V \int dz L$. 
From (\ref{sect3.2:areanoisland}, \ref{sect3.2:timenoisland},\ref{dvdzmax1},\ref{dvdzmax2}) and $v'(\bar{z}_{\max})=-\infty$, we have 
 \begin{eqnarray}\label{dAdtderive}
\lim_{t\to \infty} \frac{d A_{\text{N}}}{d t} &=&\lim_{t\to \infty} \frac{d A_{\text{N}}/d z_{\max}}{d t/d z_{\max}}= V \frac{\frac{1}{\bar{z}_{\max}^{d-2}} \sqrt{-\frac{v'(\bar{z}_{\max})(2+f(\bar{z}_{\max})v'(\bar{z}_{\max})))}{\bar{z}_{\max}^{2}} }+ \int_{z_1}^{\bar{z}_{\max}} dz \frac{\partial L}{\partial \bar{z}_{\max}}}{-\Big( v'(\bar{z}_{\max})+\frac{1}{f(\bar{z}_{\max})}\Big)- \int_{z_1}^{\bar{z}_{\max}} dz \frac{\partial v'(z)}{\partial \bar{z}_{\max}}}\nonumber\\
&=&V \frac{\sqrt{-f(\bar{z}_{\max})}}{\bar{z}^{d-1}_{\max}}= V E_{\text{N}}(\bar{z}_{\max})
\end{eqnarray}
Substituting (\ref{barzmax}) and $f(z)=1-z^{d-1}$ into (\ref{dAdtderive}), we finally obtain the late-time growth rate of $A_{\text{N}}$ (\ref{sect3.2:entropylargetime}). 

\begin{figure}[t]
\centering
\includegraphics[width=10cm]{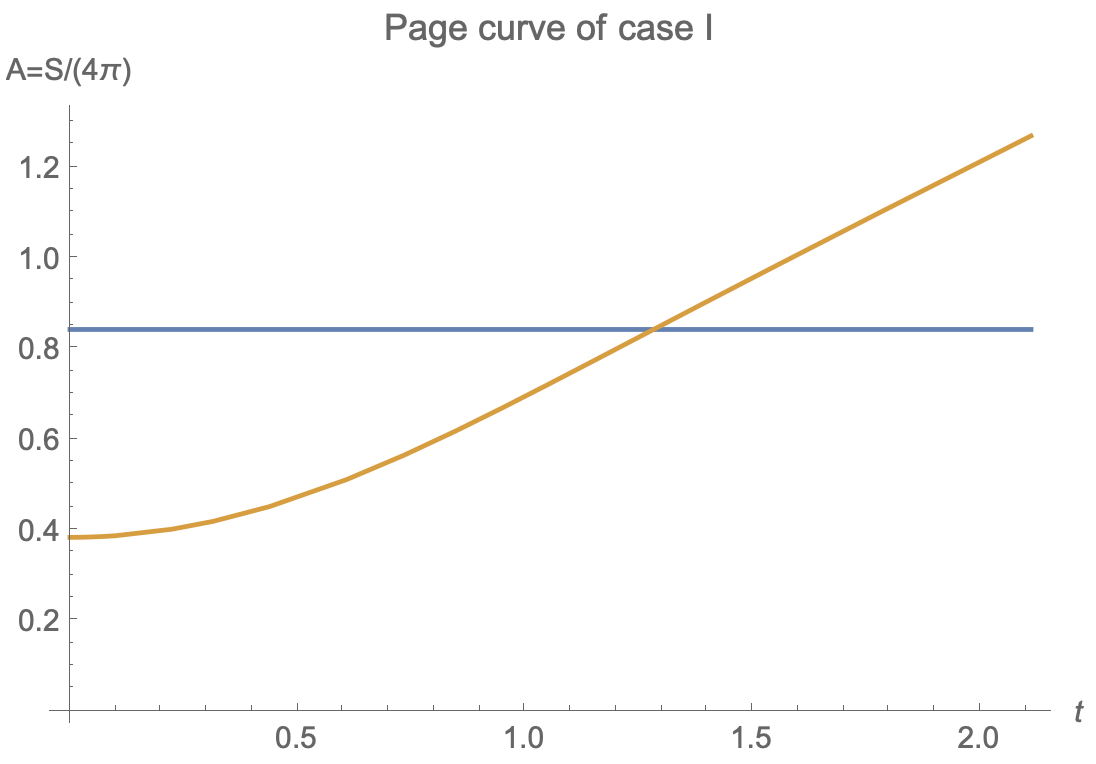}
\caption{Page curve of case I for $d=4$ and $V=1$. The orange and blue lines denote the RT surface in the no-island and island phases, respectively. The Page curve is given by the orange line before Page time, and is given by the blue after Page time.
The entanglement entropy firstly increases with time (orange line) and then becomes a constant (blue line), which recovers the Page curve of eternal black holes.}
\label{Pagecurve}
\end{figure}

We can obtain the general time dependence of $A_{\text{N}}$ by numerical calculations. The numerical method developed in \cite{Hu:2022zgy} is quite helpful in studying the HM surface. Although it is designed for codim-2 branes, it can be easily generalized to this paper's case of codim-1 branes. See appendix A for more details. We draw the Page curve in Fig. \ref{Pagecurve}, where $A$ and $t$ are half those of a two-side black hole. The entanglement entropy (orange line) increases with time at early times, and then becomes a constant (blue line) after the Page time, which reproduces the expected Page curve of the eternal black hole. 
 
 \begin{figure}[t]
\centering
\includegraphics[width=10cm]{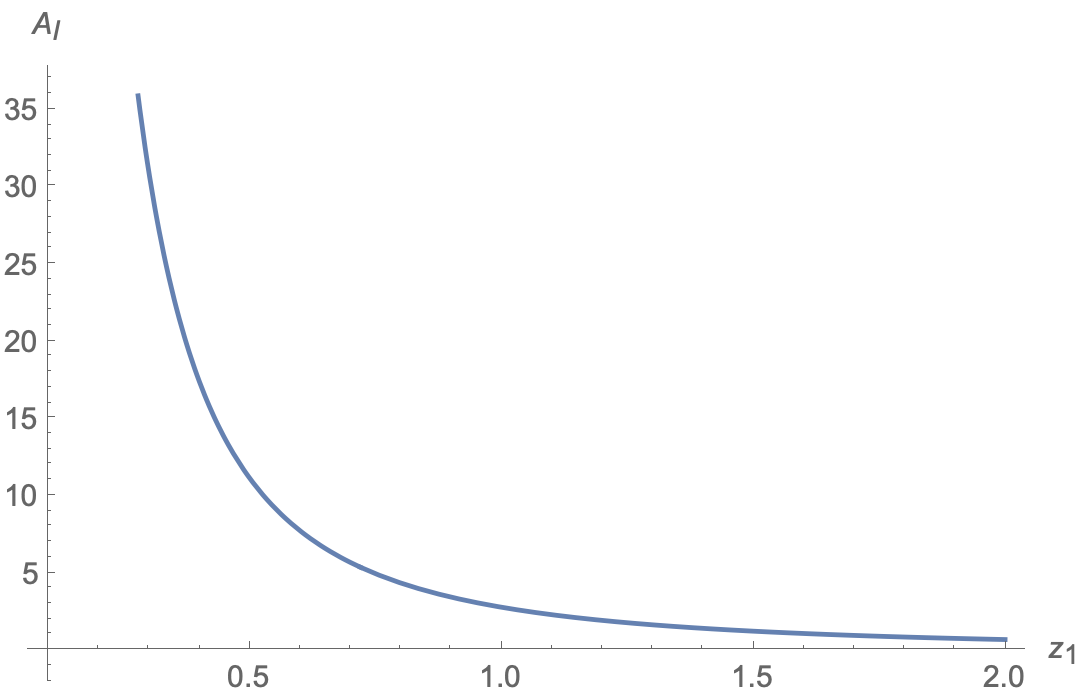}
\caption{The figure of $A_{\text{I}}(z_1)$ for an AdS space on the branes, where $A_{\text{I}}$ denotes the area of extremal surface, and $z_1$ is its endpoint on the left brane. We take the same parameters  $(\rho_1=0.5, \lambda_1=0,\rho_2=1.2, \lambda_2\approx -0.246, V=1, d=4)$ as in sect.3.1. It shows that the area of the extremal surface decreases with $z_1$ and approaches zero at the AdS horizon $z_1\to \infty$. Since the RT surface is defined as the extremal surface with minimal area, we get $A_{\text{I}}=0$ for the RT surface when the AdS black hole is replaced by an AdS space on branes. } 
\label{AIz1AdS}
\end{figure}

To end this section, let us make some comments. {\bf 1.} As shown in this section, there are non-trivial entanglement islands and Page curves in wedge holography with DGP gravity. This strongly implies the entanglement island is consistent with massless gravity.  {\bf 2}. To recover the entanglement islands in wedge holography with massless gravity on the branes, at least one of the DPG parameters $\lambda_a$ is negative. Nothing goes wrong for negative $\lambda_a$ as long as the constraints (\ref{boundDGP1},\ref{condition of kinetic energy},\ref{sect4:lambdabound}) are satisfied. We stress that our model is physically well-defined since it has positive effective Newton's constants and kinetic energy of brane bending modes, stable mass spectrum, and obeys the holographic c-theorem \cite{Miao:2022mdx}. {\bf 3.}  As discussed at the end of sect.2.2, the brane low-energy-effective theory is better approximated by Einstein's gravity for negative $\lambda_a$. That is because the massive mode is more difficult to be excited due to the more significant mass gap for negative $\lambda_a$. {\bf 4}. We get $A_{\text{I}}=0$ if the AdS black hole is replaced by an AdS space on the branes. From (\ref{sect3.1:areaisland}) with $f(z)=1$, we observe that $A_{\text{I}}\to 0$ for $z=z_a\to \infty$. See also Fig. \ref{AIz1AdS}, which shows that $A_{\text{I}}$ decreases with $z_1$ and minimizes at the AdS horizon $z_1\to \infty$. $A_{\text{I}}=0$ means entanglement entropy on the whole defect $\Sigma$ is zero for CFTs in a vacuum state, which is reasonable. {\bf 5}. One may identify the DGP term of the entropy formula (\ref{HEE}) with the second term of the island rule (\ref{islandrule}). Then it seems that $\lambda_a\sim 1/\hat{G}_N$ should be positive. However, this is not true. Note that we are studying the island rule (\ref{islandrule}) on the branes. The effective action on the brane is given by $I_{\text{eff}}= I_{\text{CFT}}+\frac{1}{16\pi G_{\text{eff N}}} \int_{Q_2} \sqrt{-h} (R_h+(d-1)(d-2)) +...$ with higher derivative corrections suppressed around the solution (\ref{metric}) \cite{Hu:2022lxl}. As a result, $\hat{G}_N$ should be identified with the effective Newton's constant on the brane instead of $1/\lambda_a$. {\bf 6}. Note that the entanglement entropy of our model is finite. It is the renormalized entanglement entropy since the branes locate at finite places instead of asymptotic infinity. Similar to Casimir energy, in principle,  the renormalized entropy can be negative. For simplicity, we do not consider this situation in this paper. {\bf 7}. The results of this paper can be generalized to cone holography \cite{Miao:2021ual}. Cone holography can be regarded as a holographic dual of the edge modes on the codim-n defect, which is a generalization of wedge holography.

\section{Page curve of case II: two black hole}

\begin{figure}[t]
\centering
\includegraphics[width=9cm]{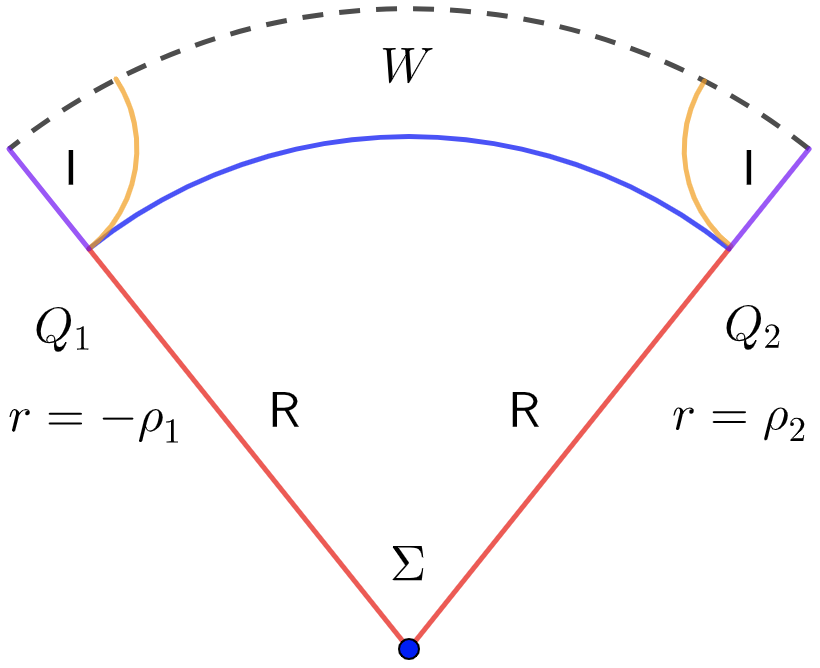}
\caption{Geometry for case II: two black holes. The red and purple lines denote the radiation $\text{R}$ and island $\text{I}$ on branes. The dotted line, blue, and orange lines indicate the horizon, RT surface in the island phase, and RT surface in the no-island phase at $t=0$, respectively. }
\label{Wedge2}
\end{figure} 

\begin{figure}[t]
\centering
\includegraphics[width=11cm]{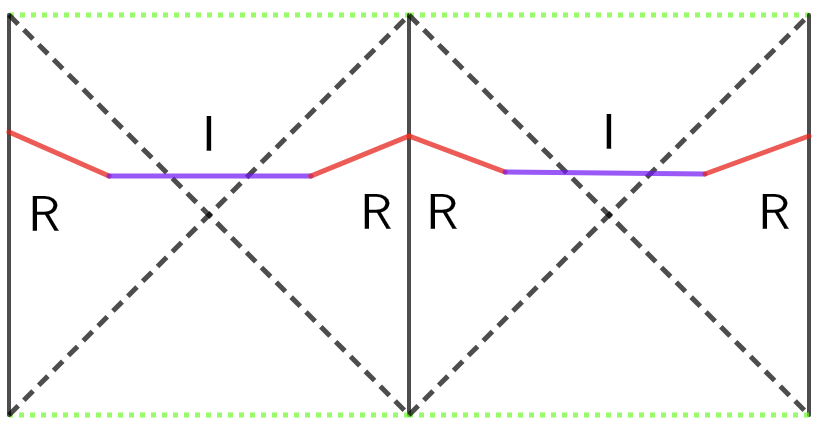}
\caption{Penrose diagram for case II: two two-side black holes. The left and right vertical lines are glued together.  The black-dotted, green-dotted, red, and purple lines denote the horizon, singularity, radiation region, and island region, respectively. }
\label{Penrose2}
\end{figure} 

In the above section, we focus on the case $G^1_{\text{eff N}} < G^2_{\text{eff N}}$ (case I), where the brane with a smaller Newton's constant can be chosen as the bath. This case describes approximately one black hole on a strong-gravity brane coupled with a bath on the weak-gravity brane. In this section, we investigate the situation $G^1_{\text{eff N}} = G^2_{\text{eff N}}$ (case II), where two black holes interact with each other on two branes of equal gravitational strength. See Fig. \ref{Wedge2} for the geometry at a time slice, and see Fig. \ref{Penrose2} for the Penrose diagram on the branes.  

Unlike case I, there is no natural way to choose the weak-gravity bath brane for case II. As a result, we have to take the two black holes on $Q_1\cup Q_2$ seriously. By symmetry and naturalness, the region near the black-hole horizon can be chosen as the island region (purple line of Fig. \ref{Wedge2}), and its complement on $Q_1\cup Q_2$ is the radiation region (red line of Fig. \ref{Wedge2}). Similar to case I, since both branes are gravitating, we adjust the radiation region $\text{R}$ (equivalently, the island region $\text{I}$, since $\partial\text{R}=\partial \text{I}$ ) to minimize the entanglement entropy of $\text{R}$ in the island phase. In this way, we fix the radiation region $\text{R}$.  Then, we can follow the usual procedure to calculate the entanglement entropy of $\text{R}$, which is given by the Hartman-Maldacena (HM) surface (orange curve of Fig.\ref{Wedge2}) at early times and given by RT surface in the island phase (blue curve of Fig.\ref{Wedge2}) at late times.  

In case II, the island region (purple line of Fig.\ref{Wedge2}) and the radiation region (red line of Fig.\ref{Wedge2}) constitute the whole space. Naturally the ``island" of case II is not one component of the radiation but that of the black hole. In other words, the ``island" of case II does not lie in the entanglement wedge of the radiation. This differs from the case with a non-gravitational bath or case I with a weak gravitational bath. As a result, the island rule (\ref{islandrule}) should be modified as
 \begin{eqnarray}\label{sect3:islandruleCaseII}
S_{\text{EE}}(\text{R})=S_{\text{EE}}(\text{I})=\text{min} \Big\{ \text{ext} \Big( S_{\text{QFT}}(\text{R})+ \frac{A(\partial \text{I})}{4 \hat{G}_N }\Big) \Big\},
\end{eqnarray}
where $\partial \text{R}=\partial \text{I}$ and $S_{\text{QFT}}(\text{R})=S_{\text{QFT}}(\text{I})$. Interestingly, (\ref{sect3:islandruleCaseII}) is  the usual formula for generalized entropy before the island theory is developed. Remarkably, this formula can give the Page curve. We give a holographic derivation of the Page curve for case II below. Since the method is the same as that of sect.3, we show only some key results below.

\begin{figure}[t]
\centering
\includegraphics[width=10cm]{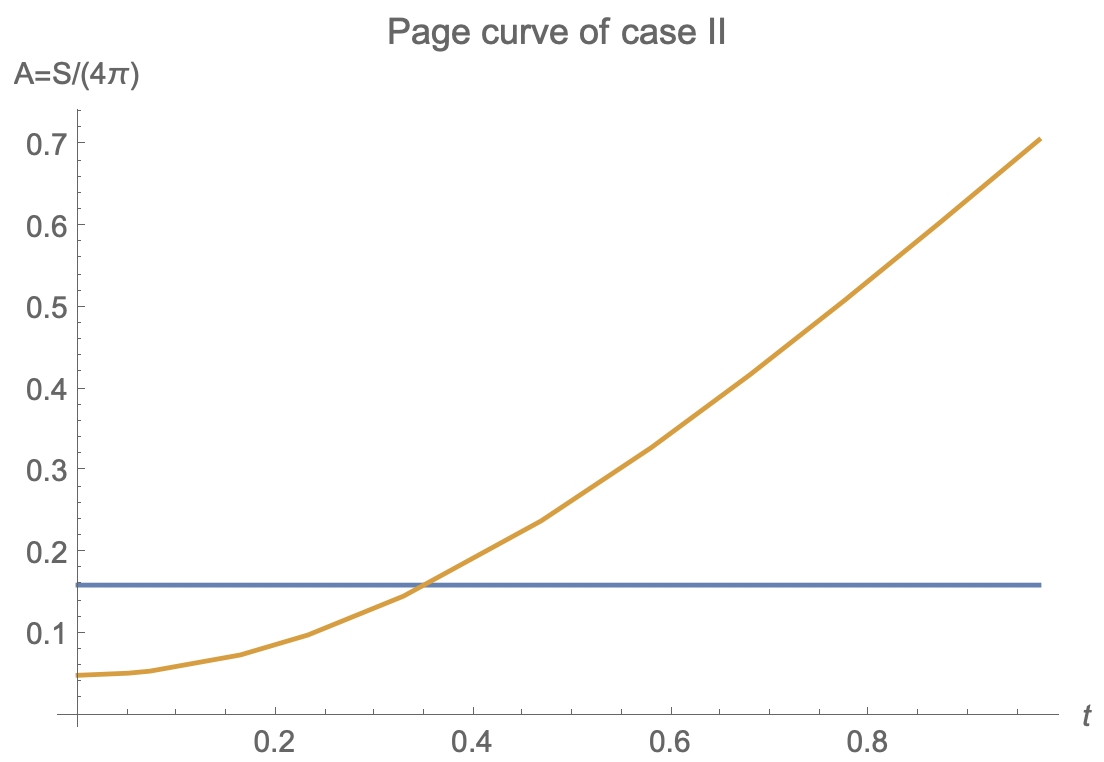}
\caption{Page curve of case II for $d=4$ and $V=1$. The orange and blue lines denote the RT surface in the no-island and island phases. The Page curve is given by the orange line before Page time, and is given by the blue after Page time. The entanglement entropy firstly increases with time (orange line) and then becomes a constant (blue line), which recovers the Page curve of the eternal black hole.}
\label{Pagecurve2}
\end{figure}

We choose the following parameters 
 \begin{eqnarray}\label{sect4:parameters }
\rho_1=\rho_2=0.5,\ \lambda_1=\lambda_2\approx -0.182,\ V=1,\ d=4,
\end{eqnarray}
which obeys the constraints (\ref{boundDGP1},\ref{condition of kinetic energy},\ref{sect4:lambdabound})
 \begin{eqnarray}\label{sect4:GNc2}
G^1_{\text{eff N}} = G^2_{\text{eff N}}\approx 0.248 >0, \ \ B \approx 0.179 >0, \ \ \lambda_1=\lambda_2 > \lambda_{\text{HEE}}\approx -0.188.
\end{eqnarray}
Following the approach of sect.3, we numerically derive the RT surface ending at $z_a\approx 0.886$ on the two branes, where $z\ge z_a$ corresponds to the island region (purple line of Fig.\ref{Wedge2}). Thus, there exist non-vanishing entanglement islands. We also obtain the area of HM surface at $t=0$, the area of RT surface in the island phase, and the black hole area with corrections from the DGP gravity as follows 
\begin{eqnarray}\label{sect4:AIANABH}
A_{\text{N}}\approx 0.049 < A_{\text{I}}\approx 0.160 < A_{\text{BH}}\approx 0.161.
\end{eqnarray}
Because $A_{\text{N}}< A_{\text{I}}$, the no-island phase dominates at the early times. As the black hole evolves, $A_{\text{N}}$ grows over time. In the large time limit, we get
\begin{eqnarray}\label{sect4:ANlarget}
\lim_{t\to \infty} \frac{d A_{\text{N}}}{dt}=V,
\end{eqnarray}
which is twice of (\ref{sect3.2:entropylargetime}) since there are two HM surfaces now. Please see the orange lines of Fig.\ref{Wedge2}. Since $A_{\text{N}}\sim t > A_{\text{I}}$ at late times, the island phase becomes dominated later, which produces the Page curve of the eternal black hole.  See Fig.\ref{Pagecurve2} for the Page curve of case II.  Finally, we want to mention that the parameters of this section also give $A_{\text{I}}=0$ if the AdS black hole is replaced by an AdS space on the branes. 

\section{Discussions on massless-island puzzle}

In sect.3 and sect.4, we show that the massless entanglement island exists in wedge holography with suitable DGP terms. And the Page curve of eternal black holes can be recovered. In this section, we discuss the puzzle of massless islands raised in \cite{Geng:2021hlu} and argue that the entanglement island is consistent with massless gravity.  

Let us first give a brief review of the massless-island puzzle \cite{Geng:2021hlu}. Following \cite{Geng:2021hlu}, we take the Karch-Randall braneworld with non-gravitational baths to illustrate the main ideas. See Fig. \ref{doubleholography} for the geometry at a constant time slice. Let us consider the late-time evolution of the black hole, where the island phase dominates, and the RT surface is given by the blue line of Fig. \ref{doubleholography}. According to the entanglement wedge reconstruction \cite{Dong:2016eik,Almheiri:2014lwa}, operators in the island region $\text{I}$ and its complement $\bar{\text{I}}$ on the brane can be reconstructed by operators in the radiation region $\text{R}$ and its complement $\bar{\text{R}}$ on the AdS boundary, respectively. Now let us focus on the d-dimensional system on the brane $Q$ and AdS boundary $M$. 
On one hand, we have a QFT system without gravity on the AdS boundary $M$. The CFT operators of $\text{R}$ commute with those of $\bar{\text{R}}$ since they are space-like separated
\begin{eqnarray}\label{sect5:RbarRcommute}
[O_{\text{R}},\ O_{\bar{\text{R}}}]=0,
\end{eqnarray}
where $O_A$ denote operators defined in the region $A$. As a result, according to the entanglement wedge reconstruction \cite{Dong:2016eik,Almheiri:2014lwa}, the operators in the island $\text{I}$ dressed to the radiation $\text{R}$ commute with operators in $\bar{\text{I}}$ dressed to $\bar{\text{R}}$
\begin{eqnarray}\label{sect5:IbarIcommute}
[O_{\text{I}},\ O_{\bar{\text{I}}}]=0.
\end{eqnarray}
On the other hand, according to the gravitational Gauss's law, the action of operators in the island $\text{I}$ must be accompanied by a disturbance in the metric outside the island, i.e., $\bar{\text{I}}$. In other words, the energy fluctuation inside $\text{I}$ can be measured in the spacetime boundary of $\bar{\text{I}}$. Thus we have
\begin{eqnarray}\label{sect5:IbarIdonotcommute}
[O_{\text{I}},\ O_{\bar{\text{I}}}]\ne 0,
\end{eqnarray}
at least for some operators, which conflicts with (\ref{sect5:IbarIcommute}). It is not a problem in the Karch-Randall braneworld since the gravity on the brane is massive and Gauss's law breakdowns. As a result, (\ref{sect5:IbarIdonotcommute}) becomes invalid. However, it is problematic for massless gravity with Gauss's law. For the above reasons, \cite{Geng:2021hlu} conjectures that the entanglement island is inconsistent with the long-range gravity obeying Gauss's law. 

\begin{figure}[t]
\centering
\includegraphics[width=6cm]{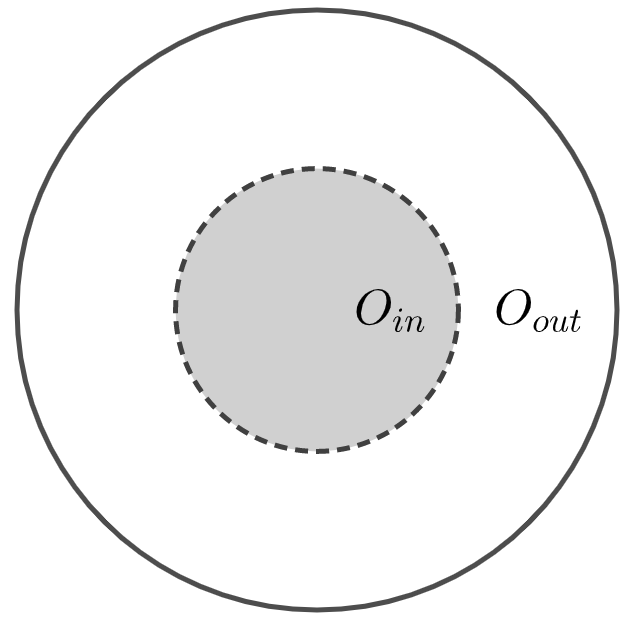}
\caption{Schematic diagram for AdS black hole, where the gray area denotes the black hole ``interior". Note that the dotted line is the RT surface derived from the generalized entropy, which is not necessarily the horizon.  $O_{in}$ and $O_{out}$ corresponds to $O_{\text{I}}$ and $O_{\bar{\text{I}}}$ in the puzzle of massless island, respectively.}
\label{BH}
\end{figure}

Now we discuss the possible resolutions to the island puzzle. To warm up,  let us study an inspiring analog of the above puzzle in AdS/CFT. Consider an AdS black hole as shown in Fig. \ref{BH}, where $O_{in}$ and $O_{out}$ are operators inside and outside the black hole region, which corresponds to $O_{\text{I}}$ and $O_{\bar{\text{I}}}$ in the above puzzle. Please note that the dotted line of Fig. \ref{BH} is the RT surface derived from the generalized entropy, which is not necessarily the horizon. For the static AdS black hole, the entanglement wedge of the whole space of CFTs is the region bounded by the RT surface and AdS boundary (white region of Fig. \ref{BH}). As a result, operators $O_{in}$ and $O_{out}$ lie outside and inside the entanglement wedges and satisfy $[O_{in},\ O_{out}]=0$ in analogy of (\ref{sect5:IbarIcommute}). On the other hand, gravity is massless in AdS/CFT. Similar to (\ref{sect5:IbarIdonotcommute}), the commutator $ [O_{in},\ O_{out}]\ne 0$ cannot vanish due to the Gauss's law. Now there is a contradiction. Of course, AdS/CFT is well-defined and cannot be wrong.  For the thermal CFT state obtained by tracing one side of the thermofield double state \cite{Maldacena:2001kr}, the puzzle can be resolved. The operator $O_{in}$ can be dressed to the other side of the black hole, which yields $[O_{in},\ O_{out}]=0$ and agrees with Gauss's law \footnote{We thank X. Dong for valuable comments on this problem.}. While for the pure thermofield double state, the problem is more subtle.  Take  $O_{in}$ and  $O_{out}$ to be the total operators on both sides, $O_{in}$  has to be dressed to the AdS boundary on one side, which leads to $ [O_{in},\ O_{out}]\ne 0$.  Recall that the black hole interior increases monotonically over time. If $O_{in}$ can be completely dressed to the fast growing black hole interior, then the puzzle can be resolved. Recently, there is an interesting paper \cite{Bahiru:2023zlc}, which finds that operators $O_{in}$  and $O_{out}$ commute, provided that the state breaks all asymptotic symmetries. If this is the case, the puzzle can be resolved too. Our key observation is that if the puzzle can be resolved in the some way in AdS/CFT, so does it in wedge holography. In fact, wedge holography is equivalent to AdS/CFT for the class of solutions (\ref{metric}) studied in this paper \cite{Miao:2020oey}, that is because they have the same effective action (\ref{effective action}).

Finally, we are ready to discuss wedge holography with DGP terms, where there is massless gravity on the branes. 
We first discuss case II with two AdS black holes on the two branes, which is very similar to the above case in AdS/CFT. Comparing Fig.(\ref{Wedge2}) with Fig.(\ref{BH}), we notice that the island region I and radiation region R of case II correspond to the grey region and white region of Fig.(\ref{BH}) in AdS/CFT.  Following the same arguments of AdS/CFT, we can resolve the potential puzzle in case II.  It should be mentioned that the puzzle \cite{Geng:2021hlu} reviewed around (\ref{sect5:RbarRcommute}-\ref{sect5:IbarIdonotcommute}) does not directly apply to case II since there are no regions of $\bar{\text{I}}$ and $\bar{\text{R}}$ in case II. Besides, the operators on the island are not described by operators in the radiation. Thus the island of case II is not the one defined in \cite{Geng:2021hlu}, which lines in the entanglement wedge of the radiation region R but is disconnected from R. In this sense, case II does not conflict with \cite{Geng:2021hlu}. Now let us turn to case I with one black hole and one weak-gravity bath. See Fig. \ref{Wedge1} for the geometry. There are several possible resolutions. 
First, unlike the case on the AdS boundary without gravity, because of Gauss's law, operators on $\text{R}$ and $\bar{\text{R}}$ do not commute anymore on the left brane with massless gravity. Thus (\ref{sect5:RbarRcommute},\ref{sect5:IbarIcommute}) breakdown and the puzzle disappears. Second, one persists in that operators on $\text{R}$ and $\bar{\text{R}}$ commute and gives up Gauss's law. According to \cite{Bahiru:2023zlc}, operators on $\text{I}$ and $\bar{\text{I}}$ commute, provided that the state breaks all asymptotic symmetries. Third, we can choose similar island and radiation regions divisions as in case II. It is a natural choice if we take into account the Hawking radiation from the black hole on the left weak-gravity brane. On the other hand, if we are interested in only the Hawking radiation from the black hole on the right strong-gravity brane, case I is a good approximation and is similar to the situation discussed in the usual double holography with non-gravitational baths.

\section{Higher derivative gravity on branes} 

In this section, we generalize the above discussions to higher derivative gravity on the branes. Higher derivative gravity is interesting in many aspects. Maybe most interestingly, the general higher curvature gravity is renormalizable \cite{Stelle}. Although it may suffer the ghost problem, one can construct a ghost-free and potentially renormalizable higher derivative gravity by choosing the parameters carefully \cite{Lu:2011zk, Biswas:2011ar, Modesto:2017sdr}. Besides, string theory predicts higher derivative corrections in gravitational action. The motivation here is to show that massless entanglement islands exist in general gravity theories. For simplicity, we focus on the following action 
\begin{eqnarray}\label{sect6:actionRR}
I&=&\int_W dx^{d+1}\sqrt{-g}\Big(R_W+d(d-1)\Big)\nonumber\\
&&+2\int_{Q} dx^d\sqrt{-h_Q}(K-T_a+\lambda_a R_{Q}+b_a \bar{R}_{Q}^2+ d_a \bar{R}_{Q ij}\bar{R}_{Q}^{\ ij}),
\end{eqnarray}
where $b_a, d_a$ are the higher derivative parameters and
\begin{eqnarray}\label{sect6:barR}
\bar{R}_Q=R_Q+d(d-1) \text{sech}^2\left(\rho _a\right),\ \bar{R}_{Q\ ij}=R_{Q\ ij}+(d-1)\text{sech}^2\left(\rho _a\right) h_{Q\ ij},
\end{eqnarray}
which vanishes for the class of solutions (\ref{metric}). As a result, at least for the solutions (\ref{metric}), the higher derivative action (\ref{sect6:actionRR}) is equal to the DGP action (\ref{action}) on-shell. However, they are different generally. In fact, $\bar{R}_{Q}^2$ and $\bar{R}_{Q ij}\bar{R}_{Q}^{\ ij}$ are ``irrelevant" higher derivative terms in the sense that they do not contribute to the Weyl anomaly \cite{Miao:2013nfa}, universal terms of entanglement entropy (logarithmic divergent term) \cite{Miao:2015iba} and correlation functions (up to three-point functions) \cite{Sen:2014nfa} for the dual CFTs. On the other hand, $\bar{R}_{Q ijkl}\bar{R}_{Q}^{\ ijkl}$ is ``relevant". However, it excludes the novel class of solutions (\ref{metric}) \footnote{Only if the bulk is a local AdS space, (\ref{metric}) is a solution to wedge holography with $\bar{R}_{Q ijkl}\bar{R}_{Q}^{\ ijkl}$ on the branes. On the other hand, the black string is no longer a solution.}. Thus, we do not consider the ``relevant" term $\bar{R}_{Q ijkl}\bar{R}_{Q}^{\ ijkl}$ in this paper and leave its study to future works. 

Similar to sect.2, we impose NBC so that there is massless gravity on the branes
\begin{eqnarray}\label{sect6:NBC}
K^{ij}-(K-T_a+\lambda_a R_{Q}) h_Q^{ij}+2 \lambda_a R^{ij}_Q+2H^{ij}=0,
\end{eqnarray}
where $\mathcal{L}=b_a \bar{R}_{Q}^2+ d_a \bar{R}_{Q ij}\bar{R}_{Q}^{\ ij}$, $\nabla_{Q\ i}$ denotes covariant derivative with respect to $h_{Q\ ij}$ and
\begin{eqnarray}\label{sect6:NBCHij}
 &&H_{ij}=P_{( i}^{\ mnl} R_{Q\ j) mnl}-2 \nabla_Q^m \nabla_Q^n P_{imnj}-\frac{1}{2} \mathcal{L} h_{Q\ ij},\\
 && P^{ijkl}=\frac{\partial \mathcal{L}}{\partial R_{Q\ ijkl}}= b_a \bar{R}_{Q} (h_Q^{ik}h_Q^{jl}-h_Q^{il}h_Q^{jk})+\frac{d_a}{2} (\bar{R}_Q^{ik}h_Q^{jl}-\bar{R}_Q^{il}h_Q^{jk}+h_Q^{ik}\bar{R}_Q^{jl}-h_Q^{il}\bar{R}_Q^{jk}).\nonumber\\
\end{eqnarray}
Note that $H_{ij}=\bar{R}_{Q ij}=\bar{R}_{Q}=P_{ijkl}=0$ for the class of solutions (\ref{metric}). As a result, the bulk metric (\ref{metric}) obeys NBC (\ref{sect6:NBC}) provided that $T_a$ and $\lambda_a$ satisfy the relation (\ref{Ta}). 

Substituting the metric (\ref{metric}) into the action (\ref{sect6:actionRR}) and integrating $r$, we get the effective action on branes
  \begin{eqnarray}\label{sect6:effective action}
I_a&=&\frac{1}{16\pi G^a_{\text{eff N}}}\int_{Q_a} \sqrt{-h} \Big( R_h+(d-1)(d-2)\Big)\nonumber\\
&&+2 \cosh(\rho_a)^{d-4}\int_{Q_a} \sqrt{-h} \Big(b_a \bar{R}_{h}^2+ d_a \bar{R}_{h\ ij}\bar{R}_{h}^{\ ij} \Big),
 \end{eqnarray} 
where $G^a_{\text{eff N}}$ denotes the effective Newton's constant (\ref{effective Newton constant}) on $Q_a$, $\bar{R}_{h}=R_h+d(d-1)$ and $ \bar{R}_{h\ ij}=R_{h\ ij}+(d-1)h_{ij}$. We require that the CFTs dual to the effective theory (\ref{sect6:effective action}) have positive central charges and no negative energy fluxes appear in scattering processes \cite{Hofman:2008ar,Buchel:2009sk}.  This yields $G^a_{\text{eff N}}>0$ but does not restrain $b_a$ and $d_a$\cite{Miao:2013nfa,Sen:2014nfa}. Usually, one treats the higher derivative terms as small corrections. Thus we focus on the case
  \begin{eqnarray}\label{sect6:conditionbd}
|b_a| <1,\ \ |d_a| <1.
 \end{eqnarray} 

It should be mentioned that the above higher-derivative model includes massless gravity on the brane. The reasons are as follows. First, since (\ref{metric}) is a solution, the induced metric on the brane obeys Einstein equations (\ref{EinsteinEQ}). Thus, it is clear that there is a massless mode. Second, the effective theory (\ref{sect6:effective action}) on the brane is a higher derivative gravity, which generally includes a massless graviton and a massive graviton. Usually, the massive mode is a ghost, which can be deleted by fine-tuning the parameters. See critical gravity \cite{Lu:2011zk, Deser:2011xc}, and the higher derivative gravity from ghost-free multi-metric gravity \cite{Hassan:2013pca, Hu:2022lxl} for examples. 

Now let us discuss the entanglement entropy of Hawking radiation. From the action (\ref{sect6:actionRR}), we can derive the holographic entanglement entropy \cite{Dong:2013qoa,Camps:2013zua}
\begin{eqnarray}\label{sect6:HEE}
S_{\text{HEE}}= 4\pi \int_{\Gamma} dx^{d-1} \sqrt{\gamma}+8\pi  \int_{\partial \Gamma} dx^{d-2} \sqrt{\sigma}\left( \lambda_a +2b_a \bar{R}_Q+ d_a(\bar{R}^{\ \alpha}_{Q\ \alpha}-\frac{1}{2} K_{\alpha} K^{\alpha} ) \right),
\end{eqnarray}
where $\Gamma$ denotes RT surface, $\partial \Gamma=\Gamma\cap Q$ is the intersection of the RT surface and the branes, $K_{\alpha}$ denote the trace of extrinsic curvatures of $\partial \Gamma$, as viewed from the brane geometry, and $\alpha$ are the directions normal to  $\partial \Gamma$ on the branes.  For the black string geometry (\ref{sect3:BHmetric}), we derive the extrinsic curvature at a constant time slice $t=\text{costant}$ and $z=z_a$ on the branes
\begin{eqnarray}\label{sect6:KK}
 K_{\alpha} K^{\alpha}=(d-2)^2 f(z_a) \text{sech}^2(\rho_a),
\end{eqnarray}
which is non-zero generally. Here $ f(z_a) =1-z_a^{d-1}$ and $z_a$ is the endpoint of the bulk RT surface on the brane $Q_a$.  Following the approach of sect.3.1, we obtain the area functional of the RT surface in the island phase
\begin{eqnarray}\label{sect6:areaisland}
A_{\text{I}}=\frac{S_{\text{HEE}}}{4\pi}&=&V\int_{-\rho_1}^{\rho_2} dr\frac{\cosh^{d-2}(r)}{z^{d-2}} \sqrt{1+\frac{\cosh^{2}(r) z'^2}{z^{2}f(z)}}\nonumber\\
&&+V\sum_{a=1}^2 \Big( \frac{2\lambda_a \cosh^{d-2}(\rho_a)}{z^{d-2}_a}- d_a (d-2)^2 f(z_a)\frac{ \cosh^{d-4}(\rho_a)}{z^{d-2}_a} \Big),
\end{eqnarray}
where we have used (\ref{sect6:KK}) and $\bar{R}_{Q ij}=\bar{R}_{Q}=0$ for the black string.  Note that only the higher derivative term $\bar{R}_{Q ij}\bar{R}_{Q}^{\ ij}$ contributes to the area (\ref{sect6:areaisland}). Consider the variation of the above area functional, we derive the NBC on the branes
\begin{eqnarray}\label{sect6:NBCisland}
\frac{(-)^az_a'}{f(z_a)\sqrt{1+\frac{\cosh^{2}(\rho_a) z_a'^2}{z_a^{2}f(z_a)}}}=\frac{2\lambda_a(d-2)z_a}{\cosh^2(\rho_a)}-\frac{d_a (d-2)^2  \left((d-2)+z_a^{d-1} \right) z_a}{\cosh ^4\left(\rho _a\right)}.
\end{eqnarray}

Following the discussions of sect.3.1, we observe that positive $d_a$ can yield a non-trivial RT surface outside the horizon. The reasons are as follows. First, the boundary term of the area functional (\ref{sect6:areaisland}) increases with $z_a$ for positive $d_a$, while the bulk term of (\ref{sect6:areaisland}) decreases with $z$ and takes minimal value at $z=1$. As a result, the total area functional (\ref{sect6:areaisland}) could minimize outside the horizon $z<1$. Second, from NBC (\ref{sect6:NBCisland}), we note that $z'_a\ne 0$, which gets rid of the no-go theorem of \cite{Geng:2020fxl} based on $z'_a=0$. Thus there could be massless entanglement islands in wedge holography with higher derivative gravity on the branes.

\begin{figure}[t]
\centering
\includegraphics[width=10cm]{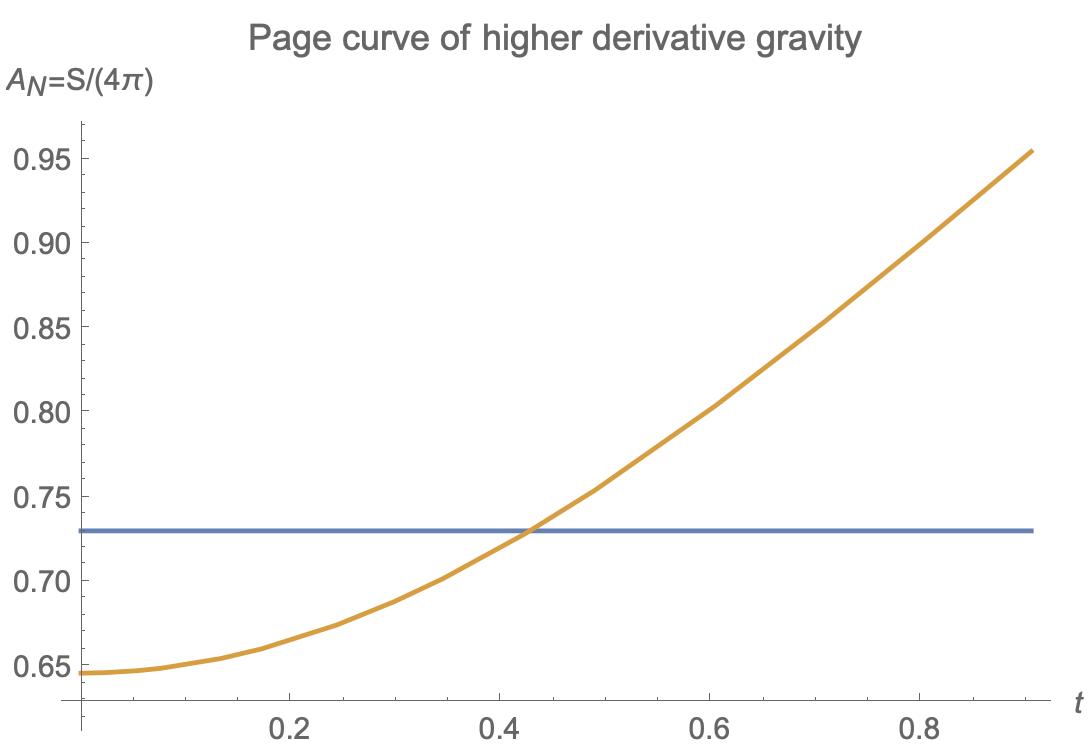}
\caption{Page curve of case I with higher derivative gravity on the branes. We have set $d=4$ and $V=1$. The orange and blue lines denote the RT surface in the no-island and island phases. The Page curve is given by the orange line before Page time, and is given by the blue after Page time. The entanglement entropy firstly increases with time (orange line) and then becomes a constant (blue line), which recovers the Page curve of the eternal black hole.}
\label{Pagecurve3}
\end{figure}

Now we are ready to study the Page curve in wedge holography with higher derivative gravity on the branes. For simplicity, we focus on case I and choose the left weak-gravity brane as the bath. Besides, we remove the DGP terms and consider only $\bar{R}_{Q ij}\bar{R}_{Q}^{\ ij}$ on the right brane. Without loss of generality, let us take the following parameters
\begin{eqnarray}\label{sect6:parameters }
d=4,\ V=1, \rho_1=0.6, \ \rho_2=0.1,\  \lambda_a=0, \ d_1=0, \ d_2\approx 0.107678 \approx 0.108,
\end{eqnarray}
which yields  
\begin{eqnarray}\label{sect6:G1G2}
0<G^1_{\text{eff N}} \approx 0.029 <  G^2_{\text{eff N}}\approx 0.198, \ \ B\approx 1.910>0.
\end{eqnarray}
Following the approaches of sect.3, we obtain the RT surface in the island phase, which starts at $z_1\approx 0.900$ on the left brane and ends on $z_2\approx 0.705$ on the right brane.  Thus the radiation region (red line of Fig.\ref{Wedge1}) locates at $z\ge z_1\approx 0.900$, and the island region (purple line of Fig.\ref{Wedge1})  locates at $z\ge z_2 \approx 0.705$. Then, we numerically derive various areas
\begin{eqnarray}\label{sect6:ANAIABH}
A_{\text{N}}\approx 0.646<A_{\text{I}}\approx 0.730< A_{\text{BH}} \approx 0.778,
\end{eqnarray}
which implies that the no-island phase dominates at the beginning $t=0$.  At late enough times, the time-growth rate of $A_{\text{N}}$ approaches a universal constant $\lim_{t\to \infty} d A_{\text{N}}/dt=V/2$. Since $A_{\text{N}}\sim t > A_{\text{I}}$ in the late times, the island phase dominates later, which recovers the Page curve of the eternal black hole. To end this section, we draw the Page curve in Fig.\ref{Pagecurve3}. Similar to sect.3, this section's higher derivative model also gives $A_{\text{I}}=0$ if the AdS black hole is replaced by an AdS space on the branes. It means that the entanglement entropy of the whole space is zero for the CFTs on the defect in a vacuum.  This is reasonable and can be regarded as a test of our model. Recall that the entanglement entropy of this paper is the renormalized entropy. Similar to Casimir energy, in principle, the renormalized entanglement entropy can be negative as long as it is bounded from below.  For simplicity, we focus on the case $A_{\text{I}}\ge 0$ in this paper.

\section{Conclusions and Discussions}

This paper investigates the entanglement island and Page curve in wedge holography with DGP gravity and higher derivative gravity on the branes. We work out the effective action for one novel class of solutions and find that the mass spectrum obeys the Breitenlohner-Freedman bound. Interestingly, the effective action and mass spectrum show that there is massless gravity on the brane. By studying the effective Newton's constant, brane bending modes and HEE, we get several lower bounds for the DGP parameters. Remarkably, there are non-trivial entanglement islands outside the horizon in wedge holography with suitable DGP gravity or higher derivative gravity on the branes. We study two cases. In case I, there is one black hole on the strong-gravity brane and a bath on the weak-gravity brane; In case II, there are two black holes on the two branes with equal gravitational strength. We find non-vanishing entanglement islands and recover the Page curve in both cases. Finally, we study an inspiring analog of the island puzzle in AdS/CFT and discuss its possible resolutions. We argue that if the contradiction can be resolved in AdS/CFT, so does it in wedge holography. Our results strongly imply that the entanglement islands exist in massless gravity theories.

There are many significant problems to explore. First, \cite{Geng:2020fxl, Geng:2022fui} prove the absence of entanglement islands in the black string geometry in the initial theory of wedge holography \cite{Akal:2020wfl}. We show that the island can be recovered in wedge holography with suitable DGP or higher derivative gravity on the branes. This raises the question if the spacetime studied in \cite{Geng:2020fxl, Geng:2022fui} is too particular. Does the entanglement island exist in more general spacetime in the initial model of wedge holography? It is a significant problem worth studying. Second, this paper only discusses the effects of curvature terms on the branes. It is interesting to see what happens when one adds appropriate matter fields on the branes. Third, we focus on the Page curve of eternal black holes. It is interesting to generalize the discussions to evaporating black holes. See some interesting progress in \cite{Emparan:2023dxm}. Fourth, there is also a massless gravitational mode on the branes of cone holography \cite{Miao:2021ual}, which generalizes wedge holography to codim-n defects. It is interesting to generalize the results of this paper to cone holography.  
Fifth, we focus on the doubly holographic model in this paper. It is a fundamental and non-trivial problem to study the entanglement islands directly in four-dimensional Einstein gravity. We hope these problems can be addressed in the future.

\section*{Acknowledgements}

We thank T. Takayanagi, X. Dong, Y. Pang, H. J. Wang, D. q. Li and Z. Q. Cui for valuable comments and discussions. This work is supported by the National Natural Science Foundation of China (No.12275366 and No.11905297).

\appendix

\section{Numerical calculation for the no-island phase}

In this appendix, we numerically calculate the time evolution of the area of the RT surface in the no-island phase. We take the same parameters as in sect.3.1. Thus, the RT surface starts at $z_1=0.95$ on the left brane and ends on the horizon $z=1$ at the beginning time $t=0$ and then passes the horizon at $t>0$. We find that $r$ approaches zero, and the area of the RT surface increases linearly with time in the late times. 

\begin{figure}[t]
\centering
\includegraphics[width=7.7cm]{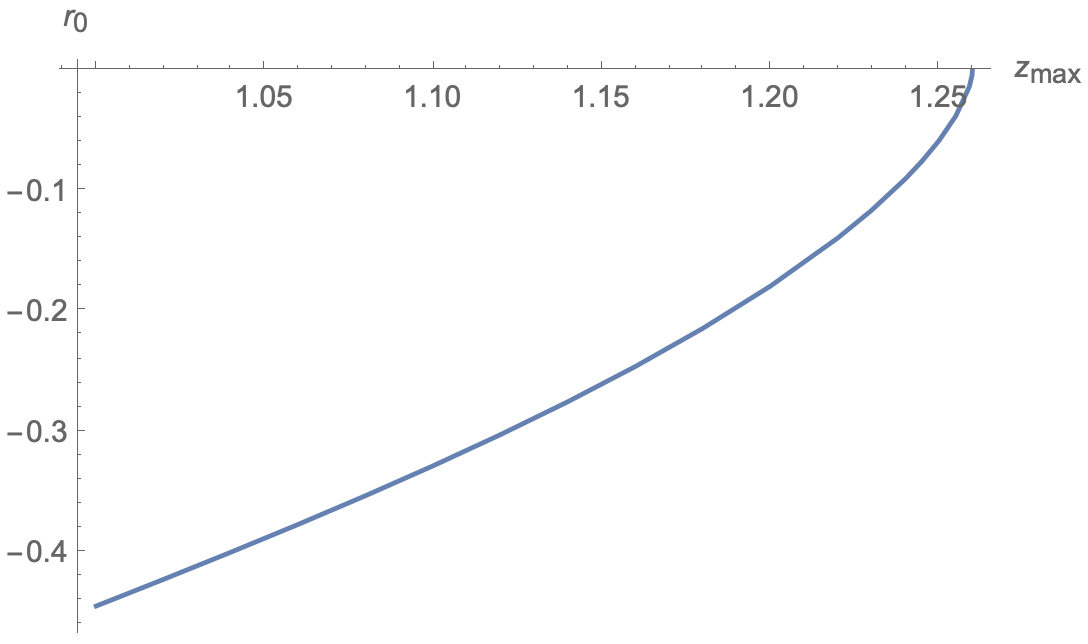}
\includegraphics[width=7.3cm]{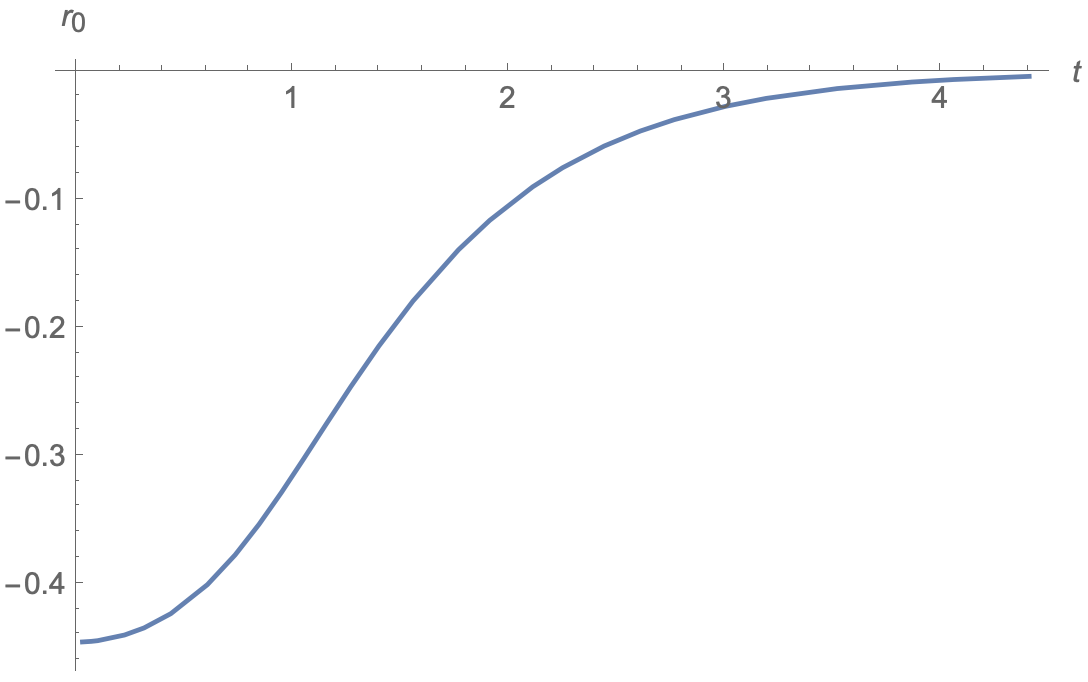}
\caption{Left: function $r_0(z_{\text{max}})$; Right: function $r_0(t)$. Note that $ z_{\text{max}}=1$ corresponds to $t=0$ and $ z_{\text{max}}=2^{\frac{1}{d-1}}\approx 1.260$ corresponds to $t\to \infty$. It shows that $r_0=r(z_{\max})$ approaches zero in the large time limit. }
\label{r0time}
\end{figure} 

\begin{figure}[t]
\centering
\includegraphics[width=7.6cm]{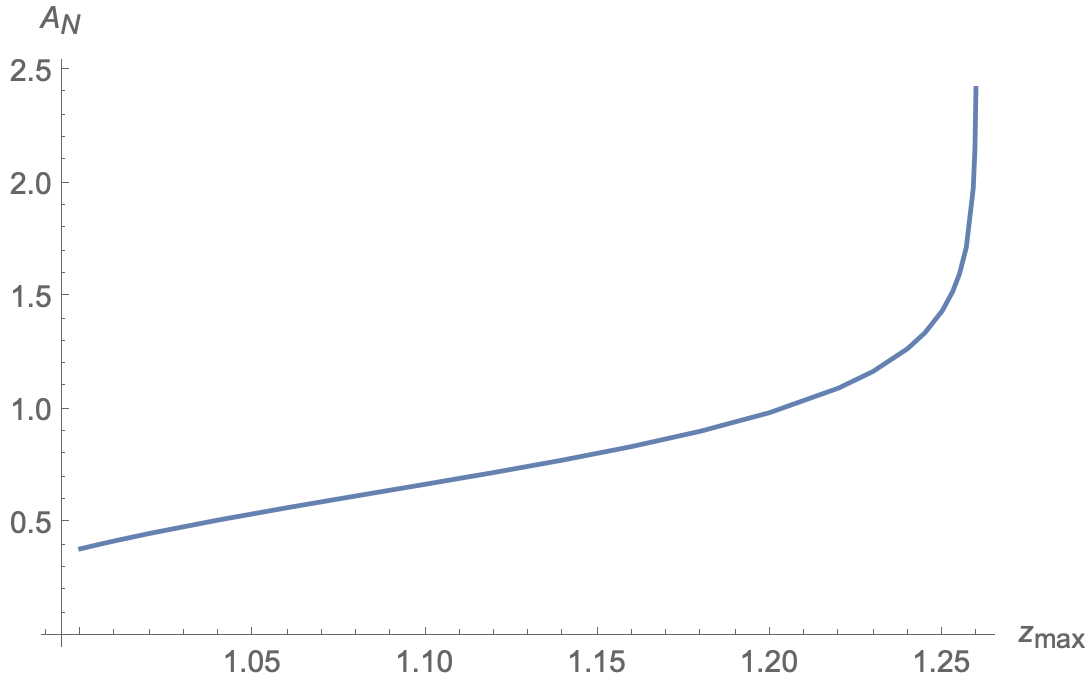}
\includegraphics[width=7.6cm]{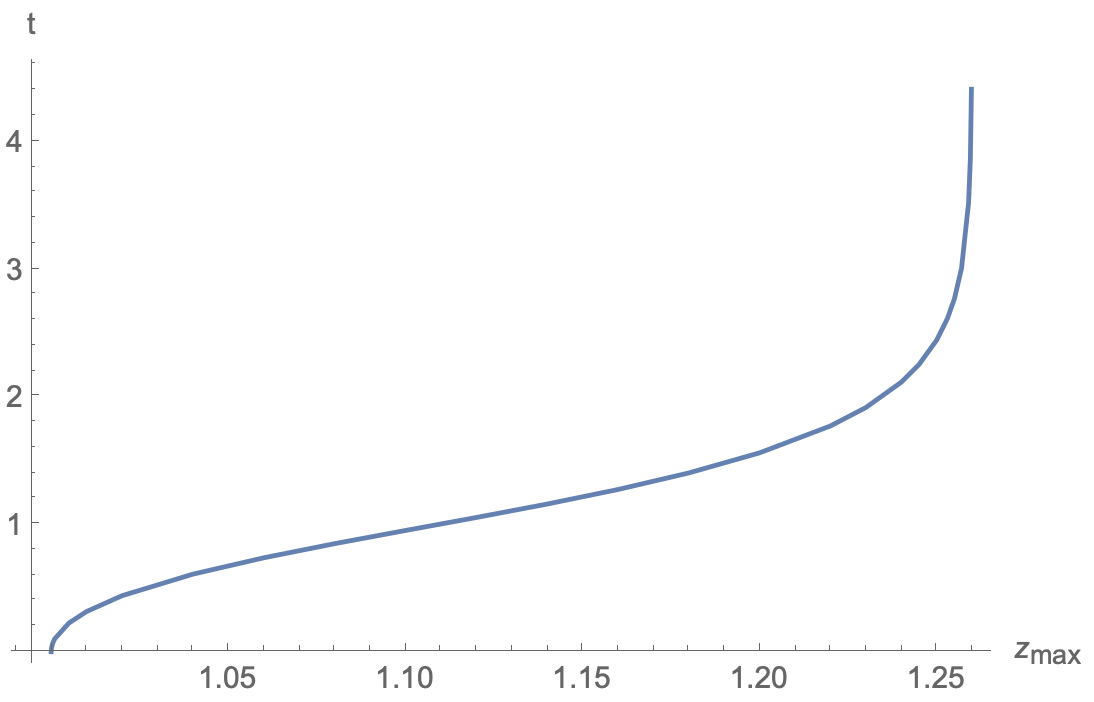}
\caption{Left: function $A_{\text{N}}(z_{\text{max}})$; Right: function $t(z_{\text{max}})$. The figures show that $A_{\text{N}}$ and $t$ increase with $z_{\text{max}}$. The right figure shows that $ z_{\text{max}}=1$ corresponds to $t=0$ and $ z_{\text{max}}=2^{\frac{1}{d-1}}\approx 1.260$ corresponds to $t\to \infty$. }
\label{ANtimezmax}
\end{figure} 

\begin{figure}[t]
\centering
\includegraphics[width=9cm]{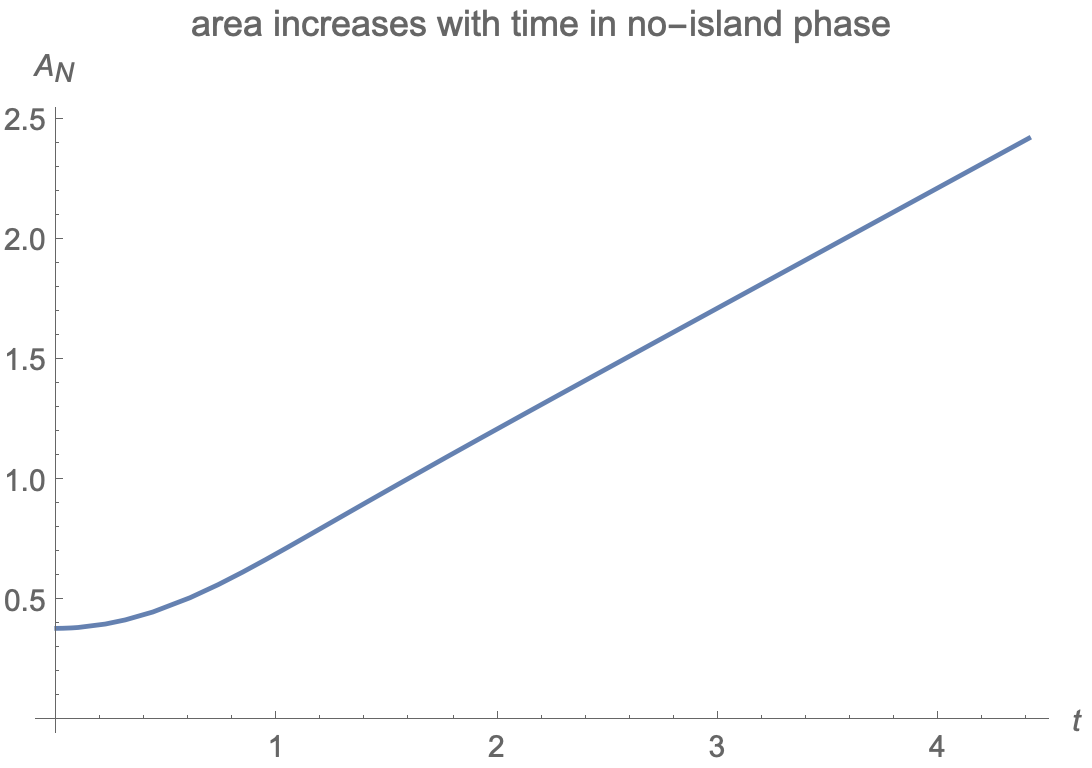}
\caption{The area of RT surface increases with time in the no-island phase. In this large time limit, we have $\lim_{t\to \infty} dA_{\text{N}}/dt=1/2$, where we have set $V=1$ for simplicity. }
\label{ANtime}
\end{figure}

Note that the area functional (\ref{sect3.2:areanoisland}), i.e., $A_{\text{N}}=V \int_{z_1}^{z_{\text{max}}} dz L(r, r', v',z)$, does not include $v(z)$ exactly. Thus we can derive a conserved quantity
 \begin{eqnarray}\label{appA:conserved quantity}
E&=&-\frac{\partial L}{\partial v'}=\frac{z^{-d} \cosh ^d(r) \left(1+f(z) v'\right)}{\sqrt{\frac{\cosh ^2(r) v' \left(-f(z) v'-2\right)}{z^2}+\left(r'\right)^2}}\nonumber\\
&=&\sqrt{-f(z_{\max })} \left(\frac{\cosh \left(r_0\right)}{z_{\max }}\right){}^{d-1},
\end{eqnarray}
where $E$ is a constant at a fixed time, $z_1$ is the endpoint of the RT surface on the left brane, $z_{\text{max}}\ge 1$ is the turning point of the two-side black hole \cite{Carmi:2017jqz}, and we have used $r(z_{\text{max}})=r_0$ and $v'(z_{\text{max}})=-\infty$ to derive the last equality of (\ref{appA:conserved quantity}).  From (\ref{appA:conserved quantity}), we can solve $v'(z)$ in functions of $r'(z)$ and $r(z)$. Substituting $v'(z)$ into the Euler-Langrangian equation derived from  (\ref{sect3.2:areanoisland}), we get the EOM of $r(z)$ decoupled with $v(z)$
 \begin{eqnarray}\label{appA: EOMr}
&&r'' \left(8 z^{2 d+1} z_{\max }^2 \cosh ^2(r) f\left(z_{\max }\right) \left(\frac{\cosh \left(r_0\right)}{z_{\max }}\right){}^{2 d}-8 f z^3 \cosh ^2\left(r_0\right) \cosh ^{2 d}(r)\right)\nonumber\\
&&-4 z^{2 d} z_{\max }^2 r' f\left(z_{\max }\right) \left(\frac{\cosh \left(r_0\right)}{z_{\max }}\right){}^{2 d} \left(z r' \left(z r' \left(z f'-2 f\right)+\sinh (2 r)\right)-4 \cosh ^2(r)\right)\nonumber\\
&&+4 z^2 r' \cosh ^2\left(r_0\right) \cosh ^{2 d-2}(r) \left(f z r' \left(2 (d-2) f z r'+d \sinh (2 r)\right)+2 (d-3) f \cosh ^2(r)-z f' \cosh ^2(r)\right)\nonumber\\
&&+8 (d-1) z \sinh (r) \cosh ^2\left(r_0\right) \cosh ^{2 d+1}(r)=0. 
\end{eqnarray}
Similarly, substituting $v'(z)$ into the area functional (\ref{sect3.2:areanoisland}) and the time (\ref{sect3.2:timenoisland}), we obtain
\begin{eqnarray}\label{appA: areanoisland}
A_{\text{N}}=V\int_{z_1}^{z_{\text{max}}} dz \left(\frac{\cosh (r)}{z}\right)^{d-2} \sqrt{\frac{2 z^2 f(z) \left(r'\right)^2+\cosh (2 r)+1}{2 z^2 \left(f(z)-f\left(z_{\max }\right) \left(\frac{z \cosh \left(r_0\right)}{z_{\max } \cosh (r)}\right){}^{2 d-2}\right)}},
\end{eqnarray}
and the time in functions of only $r(z)$. Now we have simplified the problem into solving a single differential equation (\ref{appA: EOMr}) of $r(z)$.

Solving (\ref{appA: EOMr}) perturbatively around the turning point $z=z_{\text{max}}$, we derive
\begin{eqnarray}\label{appA: rBC}
r(z)=r_0+ r_1 (z-z_{\text{max}})+ r_2  (z-z_{\text{max}})^2+O(z-z_{\text{max}})^3,
\end{eqnarray}
where 
\begin{eqnarray}\label{appA: r1}
&&r_1=\frac{\sinh \left(2 r_0\right)}{z_{\max }^d-2 z_{\max }},\\ \label{appA: r2}
&&r_2=\frac{\sinh \left(2 r_0\right) \left(2 z_{\max }^{d+1} \left((d-5) \cosh \left(2 r_0\right)-2 d-5\right)+(d+4) z_{\max }^{2 d}+24 z_{\max }^2 \cosh ^2\left(r_0\right)\right)}{6 z_{\max } \left(2 z_{\max }-z_{\max }^d\right){}^3}. \nonumber\\
\end{eqnarray}
From (\ref{appA: rBC}) we get the BC around the turning point 
\begin{eqnarray}\label{appA: rBC1}
r(z_{\text{max}}-\epsilon)=r_0+ r_1 \epsilon + r_2  \epsilon^2,\ \ r'(z_{\text{max}}-\epsilon)=r_1+ 2r_2 \epsilon, 
\end{eqnarray}
where $\epsilon$ is a small cutoff. For instance, we can choose $\epsilon=10^{-9}$.  For any given $-\rho_1< r_0 \le 0$ and $1\le z_{\text{max}} \le \bar{z}_{\max}=2^{\frac{1}{d-1}}$, we can numerically solve EOM (\ref{appA: EOMr}) with the BC (\ref{appA: rBC1}), and then derive the value of $r$ on the left brane
\begin{eqnarray}\label{appA: r on left brane}
r(z_1)=-\rho_1,
\end{eqnarray}
where we have chosen the parameters $z_1=0.95, \rho_1=-0.5$, and $d=4$ as in sect. 3.  Of course, for arbitrary inputs $r_0$ and $z_{\text{max}}$, the additional BC (\ref{appA: r on left brane}) is not satisfied generally. We can apply the shooting method to resolve this problem. For any given $1\le z_{\text{max}} \le 2^{\frac{1}{d-1}}$, we adjust the input $r_0$ so that the BC (\ref{appA: r on left brane}) is obeyed. In this way, we fix the relation between $r_0$ and $z_{\text{max}}$. See Fig. \ref{r0time} for $r_0(z_{\text{max}})$. Note that $ z_{\text{max}}=1$ corresponds to $t=0$ and $ z_{\text{max}}=2^{\frac{1}{d-1}}$ corresponds to $t\to \infty$.  

Now we have numerically solved $r(z)$ for the parameter $1\le z_{\text{max}} \le 2^{\frac{1}{d-1}}$. Substituting the solution into the area (\ref{appA: areanoisland}) and the time (\ref{sect3.2:timenoisland})\footnote{Recall that we have replaced $v'(z)$ of (\ref{sect3.2:timenoisland}) with functions of $r(z)$ solved from (\ref{appA:conserved quantity}). Now the time (\ref{sect3.2:timenoisland}) depends only on $r(z)$. }, we can derive $A_{\text{N}}(z_{\text{max}})$, $t(z_{\text{max}})$ and thus $A_{\text{N}}(t)$.  See Fig. \ref{ANtimezmax} for $A_{\text{N}}(z_{\text{max}})$ and $t(z_{\text{max}})$. See Fig. \ref{ANtime} for $A_{\text{N}}(t)$ with $V=1$, which shows that $A_{\text{N}}$ increases with time and the grown rate approaches a constant $\lim_{t\to \infty} dA_{\text{N}}/dt=1/2$ at late times, which agrees with the analytical result (\ref{sect3.2:entropylargetime}). Now we finish the numerically derivations of the time evolution of $A_{\text{N}}$.

\end{document}